\documentclass[prb,twocolumn,showpacs]{revtex4}
\usepackage{color,soul}
\usepackage{amsmath}
\usepackage{dcolumn}
\usepackage{graphicx}
\usepackage{bm}
\usepackage{amssymb}
\usepackage{subfigure}
\begin{document}

\title{Optical and Hall conductivity of the two dimensional Hubbard model: effective theory description, sign-problem-free Monte Carlo simulation and applications to the cuprate superconductors}
\author{Xinyue Liu and Tao Li}
\email{litao_phys@ruc.edu.cn}
\affiliation{School of Physics, Renmin University of China, Beijing 100872, P.R.China}

\begin{abstract}
Exact formulas for the optical conductivity and the Hall conductivity of the two dimensional Hubbard model are derived in terms of an effective theory description of the local moment fluctuation in the system. In this framework, the quantum Monte Carlo simulation of the electromagnetic response of such a strongly correlated electron system becomes sign-problem-free in many physically relevant cases. In particular, it is sign-problem-free when we assume the widely used Millis-Monien-Pines form for the phenomenological susceptibility in the effective action of the fluctuating local moment, even though these local moments are now subjected to Landau damping as a result of their coupling to the itinerant quasiparticle on the fermi surface. This is true more generally when a $\varphi^{4}$ term is included in the effective action and is thus not restricted to the Gaussian limit. Here we demonstrate the power of this framework by studying the effect of thermal fluctuation of the local moment on the optical conductivity $\sigma^{xx}(\omega)$ and the Hall conductivity $\sigma^{xy}(\omega)$ of the cuprate superconductors. Both $\sigma^{xx}(\omega)$ and $\sigma^{xy}(\omega)$ calculated are found to exhibit a two-component structure, with a Drude component at low energy and a mid-infrared component at higher energy. We find that these two components can be understood as the remnants of the corresponding spectral features contributed by the intra-band and inter-band transition in an antiferromagnetic ordered state, in which the electron band is split into the upper and the lower SDW band. For our choice of parameters, both the Drude and the mid-infrared component in $\mathrm{Im}\sigma^{xy}(\omega)$ are found to be positive and there are two sign changes between them. We find that such sign changes can be taken as the precursor of the fermi surface reconstruction, or more specifically, the emergence of electron pocket in the antiferromagnetic ordered state. Depending on the relative importance of the hole pocket and the electron pocket on the reconstructed fermi surface and the coupling strength to the local moment, the Drude component in $\mathrm{Im}\sigma^{xy}(\omega)$ can be either positive or negative. These results imply that the Hall response in the low frequency limit is rather subtle and may depends sensitively on the details of the low energy physics of the system. 
\end{abstract}

\maketitle

\section{Introduction}
An understanding of the non-Fermi liquid behavior in the normal state of the cuprate superconductors is believed to be the key toward the resolution of the mechanism of high temperature superconductivity in these strongly correlated electron systems\cite{Taillefer,Proust,Hussey}. A well known example of such non-Fermi liquid behavior is the non-Drude form of the optical absorption spectrum\cite{Basov,Marel,Heumen}. More specifically, the optical conductivity in the cuprate superconductors is found to decay much slower with frequency than that in a conventional fermi liquid metal. A two component scenario with a Drude peak in the low frequency regime and a broad mid-infrared peak extending to energy as high as the band edge seems to apply for the optical conductivity of both electron-doped and hole-doped cuprates, although it is generally believed that a single band model suffices to describe their physics. Redistribution of optical spectral weight over energy range as broad as the band width is ubiquitously observed with the variation of temperature and doping level. This is very different from the behavior of a conventional fermi liquid metal, for which the intra-band optical spectral weight is almost saturated by the narrow Drude peak.

Another example of the non-fermi liquid behavior in the normal state of the cuprate superconductors is provided by the peculiar temperature and doping dependence of their Hall response. More specifically, it is found that the Hall number $n_{\mathrm{H}}=1/R_{\mathrm{H}}$ measured in the strange metal phase of the cuprate superconductors increases almost linearly with temperature\cite{Hall1,Hall2}. In addition, $n_{\mathrm{H}}$ measured at low temperature(with superconductivity suppressed by strong magnetic field) is found to exhibit a dramatic crossover from $1+x$-like to $x$-like doping dependence around the so called pseudogap end point $x^{*}$ with the decrease of doping level $x$.\cite{Proust,Hall3} In the fermi liquid perspective, this would imply a reconstruction of the fermi surface around such a critical doping. However, no evidence for the translation symmetry breaking needed for such a fermi surface reconstruction is found around $x^{*}$. The strong contrast with the behavior of the conventional fermi liquid metal, for which the Hall number is almost temperature independent and scales linearly with the effective carrier density in the system, raises the following two questions. First, is the measured $n_{\mathrm{H}}$ still a good estimate of the effective carrier density participating in the DC transport of the system? Second, should we attribute the superconducting condensate at zero temperature totally to the itinerant spectral weight participating the DC transport just above $T_{c}$, or as proposed by some authors, if some additional spectral weight transferred from higher energy is also making significant contribution to the superconducting condensate? 

The answers to these questions are directly related to the driving mechanism of the superconductivity in the cuprate superconductors\cite{kinetic1,kinetic2,kinetic3}. More specifically, if the superconducting transition in the system is potential energy driven, as occurs in a conventional BCS superconductor, then a conserved itinerant spectral weight will be saturated at an energy scale a few times of the superconducting pairing gap. On the other hand, if the superconducting transition is kinetic energy driven as occurs in a doped Mott insultor, then spectral weight transfer over the scale of the whole band width can contribute to the transition\cite{kinetic3}. On general ground, the non-conservation of the itinerant spectral weight as implied by the non-Drude form of the optical conductivity and the peculiar temperature and doping dependence of the Hall number in the cuprate superconductors can be attributed to strong correlation effect in a doped Mott insulator, in which spectral weight redistribution occurs naturally over the energy scale of the whole band width\cite{kinetic4}. However, exactly how such a spectral weight redistribution can occur without invoking symmetry breaking transition remains elusive.

As the hallmark of a doped Mott insulator, the redistribution of electron spectral weight over the energy scale of the band width is accompanied by the emergence of fluctuating local moment, whose existence is protected by the strong correlation effect between the electrons. Such fluctuating local moment remains well-defined even when the system is far away from magnetic ordering. Indeed, RIXS measurements in the last decade find that paramagnon excitation with character similar to that of the spin wave excitation in the parent compound exists ubiquitously in the phase diagram of the cuprate superconductors\cite{Tacon,Dean}. The dual nature of electron in the cuprate superconductors as both itinerant quasiparticles and local moments poses a serious challenge to theory. However, at a phenomenological level, one can treat these two kinds of low energy excitation as independent degree of freedoms and assume a phenomenological coupling between them. The result is the so called spin-fermion model\cite{Millis,Chubukov}.

The spin-fermion model has been extensively used in the study of the high-T$_{c}$ cuprates. In particular, the theory provides a natural understanding on the origin of the d-wave pairing in the superconducting state\cite{dwave}. In addition, it is widely believed that a large number of anomalous behavior in the normal state of the cuprate superconductors can be attributed to the coupling to such fluctuating local moments\cite{Chubukov,Ueda,Lin}. To see this, let us first consider the limiting case when the fluctuating local moment condenses into a static antiferromagnetic long range order. The coupling to such an ordered background would then generate a spectral gap in the electron spectrum. Accordingly, the optical absorption spectrum will be split into the intra-band and inter-band component. The accompanying fermi surface reconstruction will also result in a change in the Hall number. Indeed, in a recent study of the Hall response in the cuprate superconductors, it is found that the calculated doping dependence of the Hall number agree well with that measured in the cuprate superconductors if we assume a nonzero spiral magnetic order in the pseudogap phase\cite{spiral,Metzner}. While the fermi surface reconstruction in the presence of such a spiral magnetic order is not consistent with the ARPES observation, the relevance of the magnetic correlation to the Hall response of the system as exposed by such a simple mean field calculation is still illuminating. A natural question that follows is whether similar transport behavior would be observed if the static spiral magnetic order assumed here is replaced by some sort of fluctuating background of the local moment. In particular, would the coupling to the fluctuating local moment generate a 'pseudogap' in the electron spectrum and a two component structure in the optical conductivity? Would the same coupling induce a dramatic suppression of the Hall number?

While such a picture is physically appealing, a reliable treatment of the effect of the local moment fluctuation in the spin-fermion model is still a challenging open problem. Unlike the electron-phonon coupling system which has an action of very similar form, the celebrated Migdal's theorem for the electron-phonon system is in general invalid for the spin-fermion model. This is not at all surprising since both the quasiparticle and the local moment in the spin-fermion model can be traced back to the same group of electrons and thus there is no natural separation in their characteristic energy scales. As a result, vertex correction effect usually can not be ignored in the study of the spin-fermion model. This is particularly the case when we consider the transport properties of the system\cite{Chubukov,Ueda,Lin}. However, a controllable treatment of such vertex correction effect is a formidable task. For example, in an attempt to address the effect of thermal spin fluctuation on the transport properties of the electron-doped cuprate superconductors, the authors of Ref.[\onlinecite{Lin}] find that if we insist on imposing the Ward identity on the current vertex function, namely adopting the conserving approximation scheme, then at the leading perturbative order the calculated conductivity would suffer an unphysical divergence in the low frequency limit\cite{Lin}. As is shown by us recently, such a divergence can be traced back to the coalescence of Green's function poles in the vertex function\cite{Liu}. Embarrassingly, while such an unphysical divergence can be cured by adopting a self-consistent Born approximation for the quasiparticle self-energy, the 'pseudogap' behavior in the electron spectrum will be removed simultaneously. We note that such a divergence is absent in the mean field treatment presented in Ref.[\onlinecite{spiral,Metzner}], since both the quasiparticle self-energy and the vertex function are calculated exactly by working in the basis of the mean field eigenstates.

An important progress in the study of the spin-fermion model occurs in 2012. In an attempt to study the quantum critical behavior around an antiferromagnetic quantum phase transition, the authors of Ref.[\onlinecite{Sachdev}] find that a suitable modification of the spin-fermion model can make it sign-problem-free for quantum Monte Carlo simulation. Such modification is usually made through doubling the flavor of the fermion in the system so that an effective anti-unitary symmetry emerges in the system. For example, the authors of Ref.[\onlinecite{Sachdev}] approximated the fermions around the two groups of hot spots on the fermi surface as two independent degree of freedoms. While such a modification may be appropriate when we discuss the quantum critical behavior around the antiferromagnetic transition, which is dominated by the fermion degree of freedom in the vicinity of the hot spots, it is in general unacceptable when we are studying the transport properties of real system, for which the contribution from the whole fermi surface should be included. 

Here we present exact formulas for the optical conductivity and the Hall conductivity of the two dimensional Hubbard model in terms of a spin-fermion-model type effective theory description of the local moment fluctuation in the system. In the formalism developed in this work, the electromagnetic(EM) response of the interacting system can be expressed as the ensemble average of the EM response kernel of a non-interacting system coupled to fluctuating local moment. We find that the simulation of the local moment fluctuation is free from the negative sign problem in either the high temperature limit or the Gaussian limit. In particular, the Monte Carlo simulation of the electromagnetic response of such a strongly correlated electron system becomes sign-problem-free when we assume the widely used Millis-Monien-Pines(MMP) form for the phenomenological susceptibility in the effective action of the fluctuating local moment, even though these local moments are subjected to Landau damping as a result of their coupling to the itinerant quasiparticle on the fermi surface. This observation remains true when a $\varphi^{4}$ term is included in the effective action and is thus not restricted to the Gaussian limit. 

We have applied our formulas to study the effect of thermal fluctuation of the local moment on the optical conductivity $\sigma^{xx}(\omega)$ and the Hall conductivity $\sigma^{xy}(\omega)$ of the cuprate superconductors. Both $\sigma^{xx}(\omega)$ and $\sigma^{xy}(\omega)$ calculated from our numerical simulation are found to exhibit a two-component structure, with a Drude component at low energy and a mid-infrared component at higher energy. We find that these two components can be understood as the remnants of the corresponding spectral features contributed by the intra-band and inter-band transition in an antiferromagnetic ordered state. We find that the Drude component in $\mathrm{Im}\sigma^{xy}(\omega)$ can be either positive or negative, depending on the relative importance of the hole pocket and the electron pocket on the reconstructed fermi surface and the coupling strength to the local moment. We find that depending on the sign of the Drude component, there are one or two sign changes between the Drude component and the mid-infrared component in $\mathrm{Im}\sigma^{xy}(\omega)$. Such sign changes can be taken as the precursor of the fermi surface reconstruction, and more specifically, the emergence of electron pocket in the antiferromagnetic ordered state. These results imply that the Hall response in the low frequency limit is very subtle and may depends sensitively on the details of the low energy physics of the system.

The paper is organized as follows. In the next section, we present a derivation of the exact formulas for the optical conductivity and the Hall conductivity of the two dimensional Hubbard model in terms of an effective theory description of the local moment fluctuation. In the third section, we present an analysis of the negative sign problem in the Monte Carlo sampling of the local moment fluctuation. The fourth section is devoted to the study of the effect of thermal spin fluctuation on the optical conductivity and the Hall conductivity of the cuprate superconductors. The last section concludes the paper and discuss the possible generalization of the current computation scheme to other situations.

\section{Effective action of the two dimensional Hubbard model and its electromagnetic response}
In this work, we will adopt the functional path integral formalism to study the transport properties of the two dimensional Hubbard model. This formalism has been adopted by Voruganti, Golubentsev, and John for the same purpose with saddle point approximation on the interaction term\cite{Hubbard,Metzner}. The Hamiltonian of the two dimensional Hubbard model reads
\begin{equation}
H=-\sum_{i,j,\sigma}t_{i,j}c^{\dagger}_{i,\sigma}c_{j,\sigma}+U\sum_{i}n_{i,\uparrow}n_{i,\downarrow}-\mu\sum_{i,\sigma}n_{i,\sigma}
\end{equation}
in which $\sigma=\uparrow,\downarrow$ denotes the spin index of the electron, $t_{i,j}$ denotes the hopping integral between site $i$ and $j$. $n_{i,\sigma}=c^{\dagger}_{i,\sigma}c_{i,\sigma}$. To derive an effective action with the form of the spin-fermion model, we rewrite the Hubbard interaction as
\begin{equation}
Un_{i,\uparrow}n_{i,\downarrow}=-\frac{2U}{3}\mathbf{s}_{i}\cdot\mathbf{s}_{i}+\frac{U}{2}(n_{i,\uparrow}+n_{i,\downarrow})
\end{equation}
where
\begin{equation}
\mathbf{s}_{i}=\frac{1}{2}\sum_{\sigma,\sigma'}c^{\dagger}_{i,\sigma}\bm{\sigma}_{\sigma,\sigma'}c_{i,\sigma'}
\end{equation} 
is the spin density operator on site $i$, $\bm{\sigma}$ is the usual spin Pauli matrix. Absorbing the last term in Eq.2 into a redefinition of the chemical potential, we have
\begin{equation}
H=-\sum_{i,j,\alpha}t_{i,j}c^{\dagger}_{i,\sigma}c_{j,\sigma}-\frac{2U}{3}\sum_{i}\mathbf{s}_{i}\cdot\mathbf{s}_{i}-\mu\sum_{i,\sigma}n_{i,\sigma}
\end{equation}

To compute the electromagnetic(EM) response of the 2D Hubbard model, we couple the electron to an electromagnetic field through the following Peierls substitution
\begin{equation}
H[\mathbf{A}]=-\sum_{i,j,\alpha}t^{A}_{i,j}c^{\dagger}_{i,\sigma}c_{j,\sigma}-\frac{2U}{3}\sum_{i}\mathbf{s}_{i}\cdot\mathbf{s}_{i}-\mu\sum_{i,\sigma}n_{i,\sigma}
\end{equation}
 in which
\begin{equation}
t^{A}_{i,j}=t_{i,j}e^{i\mathbf{A}_{i,j}(t)\cdot(\mathbf{r}_{i}-\mathbf{r}_{j})}
\end{equation}
with $\mathbf{A}_{i,j}(t)$ denoting the vector potential defined on the bond connecting site $i$ and $j$. In this paper, we will adopt the convention $\hbar=e=c=k_{B}=a=1$ for convenience, in which $a$ is the lattice constant. We will also focus on the imaginary time formalism, in which the real frequency $\omega$ should be replaced by the Matsubara frequency $i\omega_{n}$. Correspondingly, the real time $t$ should be replaced by the imaginary time $-i\tau$. To simplify notation we define $q=(\mathbf{q},i\omega_{n})$. 

The EM response of the 2D Hubbard model can be obtained from functional derivative of the free energy with respect to the EM potential 
\begin{equation}
j^{\alpha}(q)=-\frac{\delta F[\mathbf{A}]}{\delta A^{\alpha}_{-q}}
\end{equation} 
in which the free energy is determined from the partition function of the system as follows
\begin{equation}
F[\mathbf{A}]=-T\ln Z[\mathbf{A}]=-T\ln \mathrm{Tr} e^{- H[\mathbf{A}]/T}
\end{equation}
In the functional path integral formalism, the partition function $Z[\mathbf{A}]$ of the 2D Hubbard model can be represented as\cite{Nagaosa}
\begin{equation}
Z[\mathbf{A}]=\int D[\psi,\psi^{\dagger}] e^{-S[\psi,\psi^{\dagger},\mathbf{A}]}
\end{equation}
in which $\psi$ denotes the Grassmannian variable of the electron operator $c$. $S[\psi,\psi^{\dagger},\mathbf{A}]$ is the action of the system in the presence of the EM potential. It is given by
\begin{equation}
S[\psi,\psi^{\dagger},\mathbf{A}]=\int_{0}^{\beta} d\tau (\psi^{\dagger}\partial_{\tau}\psi +\mathcal{H}[\psi,\psi^{\dagger},\mathbf{A}])
\end{equation}
Here
\begin{equation} 
\psi^{\dagger}\partial_{\tau}\psi=\sum_{i,\sigma}\psi^{\dagger}_{i,\sigma}\partial_{\tau}\psi_{i,\sigma}
\end{equation} 
and
\begin{eqnarray}
\mathcal{H}[\psi,\psi^{\dagger},\mathbf{A}]&=&-\sum_{i,j,\sigma}t^{A}_{i,j}\psi^{\dagger}_{i,\sigma}\psi_{j,\sigma}\nonumber\\
&-&\frac{2U}{3}\sum_{i}\mathbf{m}_{i}\cdot\mathbf{m}_{i}-\mu\sum_{i,\sigma}\psi^{\dagger}_{i,\sigma}\psi_{i,\sigma}
\end{eqnarray}
with
\begin{equation}
\mathbf{m}_{i}=\frac{1}{2}\sum_{\sigma,\sigma'}\psi^{\dagger}_{i,\sigma}\bm{\sigma}_{\sigma,\sigma'}\psi_{i,\sigma'},
\end{equation}
To simplify notation, here and in the following we will omitted the imaginary time dependence of the field $\psi_{i,\sigma}(\tau)$ and $\mathbf{m}_{i}(\tau)$.

To proceed further, we introduce the Hubbard-Stratonovich(HS) transformation on the quartic term in the action, which reads
\begin{eqnarray}
&&\exp(\frac{2U\Delta \tau}{3} \mathbf{m}_{i}\cdot\mathbf{m}_{i} )=\left[ \frac{3\pi}{2U\Delta \tau} \right]^{-\frac{3}{2}}  \nonumber\\
&&\ \ \ \ \ \ \times\int d\bm{\phi}_{i}\exp(-\frac{2U\Delta \tau}{3} \bm{\phi}_{i}\cdot\bm{\phi}_{i} +\frac{4U\Delta \tau}{3} \bm{\phi}_{i}\cdot\mathbf{m}_{i})\nonumber\\
\end{eqnarray}
Here $\bm{\phi}_{i}$ is a vectorial HS field introduced to describe the fluctuation of the local moment. Inserting this transformation into Eq.12, we have
\begin{equation}
Z[\mathbf{A}]=\Lambda \int D[\psi,\psi^{\dagger}] D[\bm{\phi}] e^{-\int_{0}^{\beta} d\tau (\psi^{\dagger}\partial_{\tau}\psi +\mathcal{H}[\psi,\psi^{\dagger},\bm{\phi},\mathbf{A}])}
\end{equation}
in which $\Lambda$ is an unimportant constant and
\begin{eqnarray}
\mathcal{H}[\psi,\psi^{\dagger},\bm{\phi},\mathbf{A}]&=&-\sum_{i,j,\sigma}t^{A}_{i,j}\psi^{\dagger}_{i,\sigma}\psi_{j,\sigma}\nonumber\\
&-&\frac{4U}{3}\sum_{i}\bm{\phi}_{i}\cdot\mathbf{m}_{i}-\mu\sum_{i,\sigma}\psi^{\dagger}_{i,\sigma}\psi_{i,\sigma}\nonumber\\
&+&\frac{2U}{3}\sum_{i}\bm{\phi}_{i}\cdot\bm{\phi}_{i}
\end{eqnarray}

Since the action is now quadratic in the fermion field $\psi$, we can integrate it out. The result reads
\begin{equation}
Z[\mathbf{A}]=\int D[\bm{\phi}] e^{-S_{eff}[\bm{\phi},\mathbf{A}]}
\end{equation}
in which $S_{eff}[\bm{\phi},\mathbf{A}]$ is the effective action of the HS field in the presence of the EM field. It is given by
\begin{eqnarray}
S_{eff}[\bm{\phi},\mathbf{A}]&=&\int_{0}^{\beta} d\tau  \frac{2U}{3}\sum_{i}\bm{\phi}_{i}\cdot\bm{\phi}_{i}\nonumber\\
&-&\mathrm{Tr}\ln[(\partial_{\tau}-\mu-\frac{2U}{3}\bm{\phi} \cdot \bm{\sigma})-t^{A}_{i,j}]\ \nonumber\\
\end{eqnarray}
To compute the EM response of the system, we follow Ref.[\onlinecite{Hubbard,Metzner}] and rewrite $t^{A}_{i,j}$ as
\begin{equation}
t^{A}_{i,j}=t_{i,j}+u_{i,j}
\end{equation}
in which 
\begin{equation}
u_{i,j}=t_{i,j}(e^{i\mathbf{A}_{i,j}\cdot(\mathbf{r}_{i}-\mathbf{r}_{j})}-1)
\end{equation}
is to be understood as the matrix element of a matrix $u$ in real space and imaginary time(diagonal in the latter). 
Defining 
\begin{equation}
G^{-1}[\bm{\phi}]=-(\partial_{\tau}-\mu-\frac{2U}{3}\bm{\phi}\cdot\bm{\sigma})+t_{i,j}
\end{equation}
as the inverse Green's function of the electron in the presence of the fluctuation field $\bm{\phi}$ and zero EM field, we can rewrite the functional integral of $Z$ as
\begin{equation}
Z[\mathbf{A}]=\int D[\bm{\phi}] e^{-S_{eff}[\bm{\phi}]}e^{  \mathrm{Tr}\ln[I+Gu] }
\end{equation}
in which
\begin{equation}
S_{eff}[\bm{\phi}]=\int_{0}^{\beta} d\tau   \frac{2U}{3}\sum_{i}\bm{\phi}_{i}\cdot\bm{\phi}_{i}-\mathrm{Tr}\ln[-G^{-1}]  
\end{equation}
is the effective action of the HS field in the absence of the EM field.

To compute the optical conductivity and the Hall conductivity of the system at the leading order, we should expand the free energy 
\begin{equation}
F[\mathbf{A}]=-T \mathrm{ln} \left[ \int D[\bm{\phi}] e^{-S_{eff}[\bm{\phi}]}e^{  \mathrm{Tr}\ln[I+Gu] }\right]
\end{equation}
up to the third order in $\mathbf{A}$. Such an expansion is straightforward but tedious. Using the identity
\begin{eqnarray}
ln(1+x)&=&-\sum^{\infty}_{n=1}\frac{(-x)^{n}}{n} \nonumber\\
exp(x)-1&=&\sum^{\infty}_{n=1}\frac{x^{n}}{n!}
\end{eqnarray}
and the expansion
\begin{equation}
u=\sum^{\infty}_{n=1}u^{(n)}
\end{equation}
with $u^{(n)}$ the $n$-th order term of $u$ in $\mathbf{A}$, we have
\begin{eqnarray}
F[\mathbf{A}]&=&F[0]-T\sum^{\infty}_{k=1}\frac{(-1)^{k+1}}{k} \left[  \left\langle\sum^{\infty}_{l=1}\frac{1}{l!}  \left\{\sum^{\infty}_{m=1} \right. \right.\right.\nonumber\\
&&\left. \left. \left.  \frac{(-1)^{m+1}}{m} \mathrm{Tr}[(\sum^{\infty}_{n=1}Gu^{(n)})^{m}]\right\}^{l} \right\rangle\right]^{k}
\end{eqnarray}
Here $F[0]$ denotes the free energy of the system in the absence of the EM field, or more specifically
\begin{equation}
F[0]=-T \mathrm{ln} \left[ \int D[\bm{\phi}] e^{-S_{eff}[\bm{\phi}]}\right]
\end{equation}
$\langle O \rangle$ denotes the average of the observable $O$ over the distribution of the HS field $\bm{\phi}$, more specifically
\begin{equation}
\langle O \rangle=\frac{\int D[\bm{\phi}] e^{-S_{eff}[\bm{\phi}]} O }{\int D[\bm{\phi}] e^{-S_{eff}[\bm{\phi}]}}
\end{equation}

The expansion of $F[\mathbf{A}]$ to the third order in $\mathbf{A}$ are as follows.
\begin{equation}
F^{(1)}[\mathbf{A}]=0 
\end{equation}
\begin{eqnarray}
F^{(2)}[\mathbf{A}]&=&-T\left[ \left\langle \mathrm{Tr}[Gu^{(2)}] \right\rangle-\frac{1}{2}\left\langle \mathrm{Tr}[(Gu^{(1)})^{2}] \right\rangle\right.\nonumber\\ 
&+&\left. \frac{1}{2}\left\langle (\mathrm{Tr}[Gu^{(1)}])^{2} \right\rangle \right]
\end{eqnarray}
and
 \begin{eqnarray}
F^{(3)}[\mathbf{A}]&=&-T\left[  \left\langle \mathrm{Tr}[Gu^{(3)}] \right\rangle+\frac{1}{3}\left\langle \mathrm{Tr}[(Gu^{(1)})^{3}] \right\rangle   \right.\nonumber\\
&+&\frac{1}{6}\left\langle (\mathrm{Tr}[Gu^{(1)}])^{3} \right\rangle\nonumber\\
&-&\left\langle \mathrm{Tr}[Gu^{(1)}Gu^{(2)}] \right\rangle-\frac{1}{2}\left\langle \mathrm{Tr}[Gu^{(1)}] \mathrm{Tr}[(Gu^{(1)})^{2}]  \right\rangle\nonumber\\
&+&\left. \left\langle \mathrm{Tr}[Gu^{(1)}]\mathrm{Tr}[Gu^{(2)}] \right\rangle\right]
\end{eqnarray}
Here we have used the identity 
\begin{equation}
\left\langle \mathrm{Tr}[Gu^{(1)}] \right\rangle\equiv0 
\end{equation} 
This is true as a result of either the time reversal or the spatial inversion symmetry of the system. The matrix element of $u^{(n)}$ appearing in Eq.31 and Eq.32 is given by
\begin{equation}
\left[u^{(n)}\right]_{i,j}=\frac{t_{i,j}}{n!}[i\mathbf{A}_{i,j}\cdot (\mathbf{r}_{i}-\mathbf{r}_{j})]^{n}
\end{equation}
In Fourier space, the matrix element of these EM vertex becomes
\begin{eqnarray}
\left[u^{(n)}\right]_{p',p}&=&\sum_{p'=p+\sum^{n}_{i=1}q_{i}}\frac{(-1)^{n+1}}{n!}\frac{\partial^{n}\epsilon_{\mathbf{k}}}{\partial k^{\alpha_{1}}....\partial k^{\alpha_{n}}}\left |_{\frac{\mathbf{k+k'}}{2}}\right.\nonumber\\
&&\ \ \ \ \ \ \ \ \ \ \ \ \ \  \ \ \ \ \ \times A_{q_{1}}^{\alpha_{1}}....A_{q_{n}}^{\alpha_{n}}
\end{eqnarray} 
in which we use $p$ and $p'$ to denote the momentum and Matsubara frequency of the electron, namely $p=(\mathbf{k},i\nu_{n})$ and $p'=(\mathbf{k'},i\nu'_{n})$. Summation over the component index $\alpha_{1},...,\alpha_{n}$ is also implicitly assumed. For the lower order EM vertex we define the following abbreviation for convenience
\begin{eqnarray}
\left[\mathrm{v}^{\alpha}_{q}\right]_{p',p}&=&\frac{\partial\epsilon_{\mathbf{k}}}{\partial k^{\alpha}}\left |_{\frac{\mathbf{k+k'}}{2}}\right.\nonumber\\
\left[\mathrm{v}^{\alpha\beta}_{q}\right]_{p',p}&=&\frac{\partial^{2}\epsilon_{\mathbf{k}}}{\partial k^{\alpha}\partial k^{\beta}}\left |_{\frac{\mathbf{k+k'}}{2}}\right.\nonumber\\
\left[\mathrm{v}^{\alpha\beta\gamma}_{q}\right]_{p',p}&=&\frac{\partial^{3}\epsilon_{\mathbf{k}}}{\partial k^{\alpha}\partial k^{\beta}\partial k^{\gamma}}\left |_{\frac{\mathbf{k+k'}}{2}}\right.
\end{eqnarray}
in which $q=p'-p$ is the momentum and energy transfer induced by the coupling to the EM field. We note that both $\mathrm{v}^{\alpha\beta}_{q}$ and $\mathrm{v}^{\alpha\beta\gamma}_{q}$ are symmetric under the permutation of their component indices.

Now we calculate from the above expansion the EM response of the system at the leading order. To calculate the longitudinal conductivity $\sigma^{xx}$, we couple the system to a spatially uniform electric field directed in the $x$-direction of the plane. It is described by a vector potential of the form
\begin{equation}
\mathbf{A}^{E}(\mathbf{r}_{i},\tau)=(A^{E,x}_{i\omega_{n}}e^{-i\omega_{n}\tau},0,0)
\end{equation}
The corresponding electric field is given by
\begin{equation}
\mathbf{E}(\mathbf{r}_{i},\tau)=-i\frac{\partial}{\partial \tau}\mathbf{A}^{E}(\mathbf{r}_{i},\tau)=(-\omega_{n}A^{E,x}_{i\omega_{n}}e^{-i\omega_{n}\tau},0,0)
\end{equation}
or, in Fourier space
\begin{equation}
E^{x}_{i\omega_{n}}=-\omega_{n}A^{E,x}_{i\omega_{n}}
\end{equation}
It is then straightforward to find that
\begin{equation}
j^{x}(i\omega_{n})=-\frac{\partial F^{(2)}[\mathbf{A}]}{\partial A^{E,x}_{-i\omega_{n}}}=-K^{xx}(i\omega_{n})A^{E,x}_{i\omega_{n}}
\end{equation}
in which 
\begin{eqnarray}
K^{xx}(i\omega_{n})&=&T\left[\langle \mathrm{Tr}[G\mathrm{v}^{xx}_{0}]\rangle+\langle \mathrm{Tr}[G\mathrm{v}^{x}_{-i\omega_{n}}G\mathrm{v}^{x}_{i\omega_{n}}]\rangle\right.\nonumber\\
&-&\left. \langle \mathrm{Tr}[G\mathrm{v}^{x}_{-i\omega_{n}}]\mathrm{Tr}[G\mathrm{v}^{x}_{i\omega_{n}}]\rangle\right]
\end{eqnarray}
The longitudinal conductivity in real frequency is thus given by
\begin{equation}
\sigma^{xx}(\omega+i0^{+})=\frac{i}{(\omega+i0^{+})}K^{xx}(\omega+i0^{+})
\end{equation}

To describe the Hall response of the system, we should introduce both electric field and magnetic field. Here we assume that the electric field is still described by the vector potential given by Eq.37 and that the magnetic field is perpendicular to the $x-y$ plane and described by the following time-independent vector potential
\begin{equation}
\mathbf{A}^{B}(\mathbf{r}_{i},\tau)=(0,A^{B,y}_{\mathbf{q}}e^{i\mathbf{q}\cdot \mathbf{r}_{i}},0)
\end{equation}
Here the momentum $\mathbf{q}$ is directed in the $x$-direction. The corresponding magnetic field is given by
\begin{equation}
\mathbf{B}(\mathbf{r}_{i},\tau)=\nabla\times \mathbf{A}^{B}(\mathbf{r}_{i},\tau)=(0,0,iq_{x}A^{B,y}_{\mathbf{q}}e^{i\mathbf{q}\cdot \mathbf{r}_{i}})
\end{equation}
Since what concerns us is the situation when the magnetic field is spatially uniform, we should take the limit of $q_{x}\rightarrow0$ in the calculation.

By functional derivation of $F^{(3)}[\mathbf{A}]$ we have 
\begin{equation}
j^{y}(q)=-\frac{\partial F^{(3)}}{\partial A^{B,y}_{-q}}=-\kappa^{yxy}(q)A^{E,x}_{i\omega_{n}}A^{B,y}_{\mathbf{q}}=\sigma^{yx}(q)E^{x}_{i\omega_{n}}B_{\mathbf{q}}^{z}
\end{equation} 
here $q=(\mathbf{q},i\omega_{n})$,
\begin{equation}
B_{\mathbf{q}}^{z}=iq_{x}A^{B,y}_{\mathbf{q}}
\end{equation}
is the Fourier component of the magnetic field. We thus have
\begin{equation}
\sigma^{yx}(q)=\frac{1}{i\omega_{n}q_{x}}\kappa^{yxy}(q)
\end{equation}
In the  $q_{x}\rightarrow0$ limit, we have
\begin{equation}
\sigma^{yx}(\omega+i0^{+})=\frac{1}{\omega+i0^{+}}\frac{\partial}{\partial q_{x}}\kappa^{yxy}(q_{x},\omega+i0^{+})\left|_{q_{x}\rightarrow0}\right.
\end{equation}
The expression of the current response kernel $\kappa^{\alpha\beta\gamma}(q)$ is rather lengthy and it is given by
\begin{eqnarray}
\kappa^{\alpha\beta\gamma}(q)&=&T\langle \mathrm{Tr}[G\mathrm{v}^{\alpha\beta\gamma}_{0}]\rangle\nonumber\\
&+&T\langle \mathrm{Tr}[G\mathrm{v}^{\alpha}_{-q}G\mathrm{v}^{\beta}_{i\omega_{n}}G\mathrm{v}^{\gamma}_{\mathbf{q}}]\rangle\nonumber\\
&+&T\langle \mathrm{Tr}[G\mathrm{v}^{\alpha}_{-q}G\mathrm{v}^{\gamma}_{\mathbf{q}}G\mathrm{v}^{\beta}_{i\omega_{n}}]\rangle\nonumber\\
&+&T\langle \mathrm{Tr}[G\mathrm{v}^{\alpha}_{-q}G\mathrm{v}^{\beta\gamma}_{q}]\rangle\nonumber\\
&+&T\langle \mathrm{Tr}[G\mathrm{v}^{\beta}_{i\omega_{n}}G\mathrm{v}^{\alpha\gamma}_{-i\omega_{n}}]\rangle\nonumber\\
&+&T\langle \mathrm{Tr}[G\mathrm{v}^{\gamma}_{\mathbf{q}}G\mathrm{v}^{\alpha\beta}_{-\mathbf{q}}]\rangle\nonumber\\
&-&T\langle\mathrm{Tr}[G\mathrm{v}^{\alpha}_{-q}]\mathrm{Tr}[G\mathrm{v}^{\beta\gamma}_{q}]\rangle\nonumber\\
&-&T\langle\mathrm{Tr}[G\mathrm{v}^{\beta}_{i\omega_{n}}]\mathrm{Tr}[G\mathrm{v}^{\alpha\gamma}_{-i\omega_{n}}]\rangle\nonumber\\
&-&T\langle\mathrm{Tr}[G\mathrm{v}^{\gamma}_{\mathbf{q}}]\mathrm{Tr}[G\mathrm{v}^{\alpha\beta}_{-\mathbf{q}}]\rangle\nonumber\\
&+&T\langle\mathrm{Tr}[G\mathrm{v}^{\alpha}_{-q}]\mathrm{Tr}[G\mathrm{v}^{\beta}_{i\omega_{n}}]\mathrm{Tr}[G\mathrm{v}^{\gamma}_{\mathbf{q}}]\rangle\nonumber\\
&-&T[\langle\mathrm{Tr}[G\mathrm{v}^{\alpha}_{-q}G\mathrm{v}^{\beta}_{i\omega_{n}}]\mathrm{Tr}[G\mathrm{v}^{\gamma}_{\mathbf{q}}] \rangle\nonumber\\
&-&T\langle\mathrm{Tr}[G\mathrm{v}^{\alpha}_{-q}G\mathrm{v}^{\gamma}_{\mathbf{q}}]\mathrm{Tr}[G\mathrm{v}^{\beta}_{i\omega_{n}}] \rangle\nonumber\\
&-&T\langle\mathrm{Tr}[G\mathrm{v}^{\beta}_{i\omega_{n}}G\mathrm{v}^{\gamma}_{\mathbf{q}}]\mathrm{Tr}[G\mathrm{v}^{\alpha}_{-q}] \rangle
\end{eqnarray}
Here we note again that $q=(\mathbf{q},i\omega_{n})$. $\mathrm{v}^{\alpha}_{\mathbf{q}}$ and $\mathrm{v}^{\alpha\beta}_{\mathbf{q}}$ here and in the following are abbreviations of $\mathrm{v}^{\alpha}_{q=(\mathbf{q},i\omega_{n}=0)}$ and $\mathrm{v}^{\alpha\beta}_{q=(\mathbf{q},i\omega_{n}=0)}$.

\section{Quantum Monte Carlo simulation of the effective action and the negative sign problem}
Given the expression for the effective action $S_{eff}[\bm{\phi}]$, we can in principle compute the EM kernel $K^{xx}(q)$ and $\kappa^{\alpha\beta\gamma}(q)$ from Eq.41 and Eq.49 through quantum Monte Carlo sampling of the fluctuating field $\bm{\phi}$. However, the effective action $S_{eff}[\bm{\phi}]$ of $\bm{\phi}$ as obtained from the integration over the fermion field is in general not real. The quantum Monte Carlo simulation of $e^{-S_{eff}[\bm{\phi}]}$ thus in general suffers from the negative sign problem. Here we assume that the system has a general incommensurate filling so that there is no symmetry to guarantee the exact cancellation of the negative sign problem\cite{Hirsch}. An exact treatment of such negative sign problem in quantum Monte Carlo simulation is still an open problem.

Here we show that $e^{-S_{eff}[\bm{\phi}]}$ is positive definite in some important limiting cases that are of physical relevance. The first case is the high temperature limit in which $\beta\rightarrow 0$. We can now ignore the time dependence of the fluctuating field $\bm{\phi}$. In such a case, $S_{eff}[\bm{\phi}]$ is nothing but the free energy of the electron in the background of a static Zeeman field $\frac{4U}{3}\bm{\phi}_{i}$, which is obviously real. We will focus on such a limiting case in the next section.

Another interesting limiting case is when the perturbative expansion in the interaction strength $U$ converges. More specifically, since
\begin{eqnarray}
&&\mathrm{Tr}\ln[-G^{-1}[\bm{\phi}]] =\mathrm{Tr}\ln[-G_{0}^{-1}-\frac{2U}{3}\bm{\phi}_{i}\cdot \mathbf{m}_{i}] \nonumber\\
&=&\mathrm{Tr}\ln[-G_{0}^{-1}]+\mathrm{Tr}\ln[I+\frac{2U}{3}G_{0}\bm{\phi}_{i}\cdot \mathbf{m}_{i}] \nonumber\\
&=&\mathrm{Tr}\ln[-G_{0}^{-1}]-\sum^{\infty}_{n=1}\frac{(-2U/3)^{n}}{n}\mathrm{Tr}[G_{0}\bm{\phi}_{i}\cdot \mathbf{m}_{i}]^{n}\nonumber\\
\end{eqnarray}
in which $G_{0}=[-(\partial_{\tau}-\mu)+t_{i,j}]^{-1}$ is the inverse Green's function of the non-interacting system. Since $G_{0}$ is by definition real and spin rotationally invariant, all terms in the perturbative expansion are real in the imaginary time domain. In particular, to the second order in $U$ and up to an unimportant real constant we have
\begin{equation}
S_{eff}[\bm{\phi}] \approx\sum_{q}[\frac{2U}{3}-\frac{2U^{2}}{9}\chi_{0}(q)]\bm{\phi}_{q}\cdot \bm{\phi}_{-q}
\end{equation}
Here $\chi_{0}(q)$ is the bare spin susceptibility of the system and is given by
\begin{equation}
\chi_{0}(q)=\mathrm{Tr}[G_{0} \mathbf{m}_{q}G_{0}\mathbf{m}_{-q}]
\end{equation} 
in which $\mathbf{m}_{q}$ denotes the Fourier component of the spin density operator of the electron. Using the property that 
\begin{equation}
\chi_{0}(q)=\chi_{0}(-q)=(\chi_{0}(q))^{*}
\end{equation} 
we know that $S_{eff}[\bm{\phi}]$ is a real number and thus $e^{-S_{eff}[\bm{\phi}]}$ is positive-definite in the Gaussian limit.

In the strong coupling regime in which fluctuating local moment emerges in the low energy physics, the perturbative expansion in $U$ no longer converges and the negative sign problem is in general unavoidable. Here we assume that the low energy physics of the fluctuating local moment can be approximated by a sign-problem-free effective action. A well known such approximation is the spin-fermion model, in which the action of the fluctuating local moment is assumed to take the following Gaussian form\cite{Chubukov}
\begin{equation}
S_{eff}[\bm{\phi}]=\frac{1}{2}\int_{0}^{\beta} d\tau\int_{0}^{\beta} d\tau'\sum_{\mathbf{q}} \chi^{-1}(\mathbf{q},\tau-\tau') \bm{\phi}_{\mathbf{q}}(\tau) \cdot\bm{\phi}_{-\mathbf{q}}(\tau')
\end{equation}
Here $\chi(\mathbf{q},\tau-\tau')$ denotes the generalized susceptibility of the local moment. A widely adopted form for $\chi$ is the MMP susceptibility proposed by Millis, Monien and Pines(MMP) in the early 1990s\cite{Millis}. The MMP susceptibility has a Ornstein-Zernike form and reads 
\begin{equation}
\chi(\mathbf{q},i\omega_{n})=\frac{\chi(\mathbf{Q})}{1+(\mathbf{q}-\mathbf{Q})^{2}\xi^{2}+|\omega_{n}|/\omega_{sf}}
\end{equation}
in which $\chi(\mathbf{Q})$ denotes the static susceptibility at the antiferromagnetic wave vector $\mathbf{Q}=(\pi,\pi)$. $\xi$ measures the correlation length of the antiferromagnetic correlated local moment. $\omega_{sf}$ denotes the characteristic frequency of the Landau damped fluctuation of the local moment as a result of its coupling to the itinerant quasiparticles on the fermi surface. These parameters should be understood as phenomenological parameters to be determined from fitting experimental spin fluctuation spectrum. At a more microscopic level, the MMP susceptibility can be derived from the expansion of the RPA susceptibility near an antiferromagnetic quantum critical point, in which $\chi(\mathbf{Q})\propto \xi^{2}$, $\omega_{sf}\propto\xi^{-2}$.  

An important property of the MMP susceptibility is that it is real in imaginary frequency domain. This is actually true for any meaningful spin susceptibility, as can be seen from the following spectral representation of $\chi$. More specifically, we have
\begin{eqnarray}
\chi(\mathbf{q},i\omega_{n})&=&\frac{1}{2\pi}\int^{\infty}_{-\infty}d\omega' \frac{R(\mathbf{q},\omega')}{i\omega_{n}-\omega'}\nonumber\\
&=&-\frac{1}{\pi}\int_{0}^{\infty}\frac{\omega'R(\mathbf{q},\omega')}{\omega_{n}^{2}+\omega'^{2}}
\end{eqnarray}
in which $R(\mathbf{q},\omega)$ denotes the real spectral function of the spin fluctuation and it satisfy $R(\mathbf{q},\omega)=-R(\mathbf{q},-\omega)$. 
Thus the Monte Carlo sampling of the distribution $ e^{-S_{eff}[\bm{\phi}]}$ is free from the negative sign problem in imaginary time domain when the effective action has the form of Eq.54. Clearly, a $\phi^{4}$ term can be added to $S_{eff}[\bm{\phi}]$ without introducing additional negative sign problem. Thus, a sign-problem-free sampling of the local moment fluctuation is not restricted to the Gaussian regime and has much broader scope of applications. 

From the Monte Carlo sampling of $e^{-S_{eff}[\bm{\phi}]}$, we can obtain both $K^{xx}(q)$ and $\kappa^{\alpha\beta\gamma}(q)$ in imaginary frequency domain. To obtain the conductivity in the real frequency domain, a Wick rotation $i\omega\rightarrow \omega+i0^{+}$ is needed. Such a Wick rotation can be done through numerical analytic continuation. Although such a numerical procedure suffers from the ambiguity related to the exponential suppression of Boltzmann weight at large imaginary time, we note that even the EM response kernel computed in the imaginary frequency domain already contain valuable physical information that can be used to check the consistency of a given theory.

\section{Optical conductivity and Hall conductivity in the presence of thermal fluctuation of local moment}

\subsection{General considerations}  
The theory presented in Sec.II allow us to compute the EM response of the system in a numerically exact fashion through Monte Carlo sampling over the field configuration $\bm{\phi}$, provided that the distribution $e^{-S_{eff}[\bm{\phi}]}$ is positive definite. However, the lack of both the computational resource and the numerical expertise to perform the full quantum Monte Carlo simulation and the numerical analytic continuation make this goal currently beyond our reach. Here we will demonstrate the power of the formalism developed in Sec.II by studying the much simpler case in which thermal fluctuation of the local moment dominates the effective action. Such a limiting case has been considered in perturbative framework by Lin and Millis following the theory developed by Lee, Rice and Anderson(LRA) for system with fluctuating charge density wave\cite{Lin,LRA}. As argued in Ref.[\onlinecite{Lin}], this limiting case have strong physical relevance for the electron-doped cuprate superconductors, in which thermal fluctuation of antiferromagnetic correlated local moment exists in broad regions of the phase diagram. However, as a result of the difficulty in the treatment of the vertex correction in the EM response, the perturbative result obtained by Lin and Millis suffers from unphysical divergence in the low frequency limit. We will see that our formalism is free from such unphysical divergence and can produce numerical accurate prediction for the EM response of an electron system coupled to thermal fluctuation of local moment at any frequency.

Another reason for us to focus on the thermal fluctuation effect is that we can now compute the EM response of the system directly in the real frequency domain, avoiding the subtleties related to the ill-conditioned numerical analytic continuation for processing the imaginary time data generated by the full quantum Monte Carlo simulation\cite{Imada,Huang,Wang}. This is particularly the case in the study of the Hall conductivity, since $\mathrm{Im}\sigma^{xy}(\omega)$ in general can change its sign\cite{Huang,Wang}. We also note that the assumption of the dominance of thermal fluctuation will eventually break down when we consider the EM response of the system in the low frequency limit. In such a limit, the quantum nature of the local moment fluctuation will play a crucial role. For this reason, we will focus on the EM response of the system at finite frequency. We note that there are an abundance of experimental results on the optical conductivity\cite{Basov} and Hall conductivity in the optical frequency regime for the cuprate superconductors\cite{infrared0,infrared1,infrared2,RSI,infrared3,infrared4,infrared5}. However, the discussion below will be restricted at a qualitative rather than quantitative level. No attempt to fit the experimental data of the cuprate superconductors will be exercised.

\subsection{Single particle Green's function and the EM response kernel in the presence of thermal fluctuation of local moment }
Since we focus on thermal fluctuation of the local moment, we can ignore the time dependence of the field $\bm{\phi}$. As a result, the electron Green's function becomes diagonal in the Matsubara frequency. More specifically, the electron Green's function in the background of a given field configuration $\bm{\phi}$ can be written as 
\begin{equation}
G_{\mathbf{q}\sigma,\mathbf{q'}\sigma'}(i\nu_{n})=\sum_{m}\frac{\varphi^{m}_{\mathbf{q}\sigma}\ \varphi^{m*}_{\mathbf{q'}\sigma'}}{i\nu_{n}-E_{m}}
\end{equation}
in which $i\nu_{n}=(2n+1)\pi T$ is the fermion Matsubara frequency, $\varphi^{m}_{\mathbf{q}\sigma}$ is the wave function of the $m$-th eigenstate of $\mathcal{H}[\psi,\psi^{\dagger},\bm{\phi},\mathbf{A}=0]$ in the momentum representation. $E_{m}$ is its energy.

The computation of the EM response kernel $K^{xx}(q)$ and $\kappa^{\alpha\beta\gamma}(q)$ reduce to the computation of the various fermion traces appearing in Eq.41 and Eq.49. Since $G$ is now diagonal in $i\nu_{n}$, we have
\begin{eqnarray}
\mathrm{Tr}[G\mathrm{v}^{\alpha}_{q}]&=&0\nonumber\\
\mathrm{Tr}[G\mathrm{v}^{\alpha}_{i\omega_{n}}]&=&0\nonumber\\
\mathrm{Tr}[G\mathrm{v}^{\alpha\beta}_{i\omega_{n}}]&=&0\nonumber\\
\mathrm{Tr}[G\mathrm{v}^{\alpha}_{i\omega_{n}}G\mathrm{v}^{\beta}_{\mathbf{q}}]&=&0\nonumber\\
\mathrm{Tr}[G\mathrm{v}^{\alpha}_{q}]&=&0\nonumber\\
\mathrm{Tr}[G\mathrm{v}^{\alpha\beta}_{q}]&=&0
\end{eqnarray}
for $\omega_{n}\neq0$. Here $q=(\mathbf{q},i\omega_{n})$.  As a result, we have
\begin{eqnarray}
K^{xx}(i\omega_{n})=T[\langle\mathrm{Tr}[G\mathrm{v}^{xx}_{\mathbf{q}=0}]\rangle+\langle\mathrm{Tr}[G\mathrm{v}^{x}_{-i\omega_{n}}G\mathrm{v}^{x}_{i\omega_{n}}]\rangle]
\end{eqnarray}
and
\begin{eqnarray}
\kappa^{\alpha\beta\gamma}(q)&=&T\langle \mathrm{Tr}[G\mathrm{v}^{\alpha\beta\gamma}_{0}]\rangle\nonumber\\
&+&T\langle \mathrm{Tr}[G\mathrm{v}^{\alpha}_{-q}G\mathrm{v}^{\beta}_{i\omega_{n}}G\mathrm{v}^{\gamma}_{\mathbf{q}}]\rangle\nonumber\\
&+&T\langle \mathrm{Tr}[G\mathrm{v}^{\alpha}_{-q}G\mathrm{v}^{\gamma}_{\mathbf{q}}G\mathrm{v}^{\beta}_{i\omega_{n}}]\rangle\nonumber\\
&+&T\langle \mathrm{Tr}[G\mathrm{v}^{\alpha}_{-q}G\mathrm{v}^{\beta\gamma}_{q}]\rangle\nonumber\\
&+&T\langle \mathrm{Tr}[G\mathrm{v}^{\beta}_{i\omega_{n}}G\mathrm{v}^{\alpha\gamma}_{-i\omega_{n}}]\rangle\nonumber\\
&+&T\langle \mathrm{Tr}[G\mathrm{v}^{\gamma}_{\mathbf{q}}G\mathrm{v}^{\alpha\beta}_{-\mathbf{q}}]\rangle\nonumber\\
&-&T\langle\mathrm{Tr}[G\mathrm{v}^{\gamma}_{\mathbf{q}}]\mathrm{Tr}[G\mathrm{v}^{\alpha\beta}_{-\mathbf{q}}]\rangle\nonumber\\
&-&T[\langle\mathrm{Tr}[G\mathrm{v}^{\alpha}_{-q}G\mathrm{v}^{\beta}_{i\omega_{n}}]\mathrm{Tr}[G\mathrm{v}^{\gamma}_{\mathbf{q}}] \rangle
\end{eqnarray}
The above expression for $\kappa$ can be further simplified using the lattice symmetry of the system. More specifically, using the rotational symmetry of the system about the $z$-axis, we know that in the $|\mathbf{q}|\rightarrow0$ limit only the asymmetric component of $\kappa^{\alpha\beta\gamma}(q)$ with respect to $\alpha$ and $\beta$ can contribute to the Hall conductivity of the system. We can thus ignore the first, the sixth and the seventh term in Eq.60 since they are manifestly symmetric with respect to $\alpha$ and $\beta$. At the same time, the inversion symmetry of the system dictates that the fifth and the last term in Eq.60 are both even function of momentum $\mathbf{q}$ and thus do not contribute to the Hall conductivity in the $|\mathbf{q}|\rightarrow0$ limit. With these considerations in mind, we have 
\begin{equation}
\sigma^{yx}(\omega+i0^{+})=\frac{1}{\omega+i0^{+}}\frac{\partial}{\partial q_{x}}\tilde{\kappa}^{yxy}(q_{x},\omega+i0^{+})\left|_{q_{x}\rightarrow0}\right.
\end{equation}
in which $\tilde{\kappa}^{yxy}$ has the much simpler expression of
\begin{eqnarray}
\tilde{\kappa}^{\alpha\beta\gamma}(q)&=&T\langle \mathrm{Tr}[G\mathrm{v}^{\alpha}_{-q}G\mathrm{v}^{\beta}_{i\omega_{n}}G\mathrm{v}^{\gamma}_{\mathbf{q}}]\rangle\nonumber\\
&+&T\langle \mathrm{Tr}[G\mathrm{v}^{\alpha}_{-q}G\mathrm{v}^{\gamma}_{\mathbf{q}}G\mathrm{v}^{\beta}_{i\omega_{n}}]\rangle\nonumber\\
&+&T\langle \mathrm{Tr}[G\mathrm{v}^{\alpha}_{-q}G\mathrm{v}^{\beta\gamma}_{q}]\rangle
\end{eqnarray}  

The fermion traces in Eq.59 and Eq.62 can be calculated straightforwardly, with the results given by
\begin{eqnarray}
T\mathrm{Tr}[G\mathrm{v}^{\alpha\beta}_{\mathbf{q}}]&=&\sum_{m}[\mathrm{v}^{\alpha\beta}_{\mathbf{q}}]_{m,m}f(E_{m})\nonumber\\
T\mathrm{Tr}[G\mathrm{v}^{\alpha}_{-i\omega_{n}}G\mathrm{v}^{\beta}_{i\omega_{n}}]&=&\sum_{m,m'}[\mathrm{v}^{\alpha}_{0}]_{m',m}[\mathrm{v}^{\beta}_{0}]_{m,m'}\nonumber\\
&\times&\frac{f(E_{m'})-f(E_{m})}{i\omega_{n}+(E_{m'}-E_{m})}\nonumber\\
\end{eqnarray}
\begin{eqnarray}
T\mathrm{Tr}[G\mathrm{v}^{\alpha}_{-q}G\mathrm{v}^{\beta\gamma}_{q}]&=&\sum_{m,m'}[\mathrm{v}^{\alpha}_{-\mathbf{q}}]_{m',m}[\mathrm{v}^{\beta\gamma}_{\mathbf{q}}]_{m,m'}\nonumber\\
&\times&\frac{f(E_{m'})-f(E_{m})}{i\omega_{n}+(E_{m'}-E_{m})}
\end{eqnarray}
and
\begin{eqnarray}
&&T\mathrm{Tr}\left[G\mathrm{v}^{\alpha}_{-q}G\mathrm{v}^{\beta}_{i\omega_{n}}G\mathrm{v}^{\gamma}_{\mathbf{q}}\right]=\nonumber\\
&&\ \ \ \ \ \ \ \ \ \sum_{m,m',m''}\left[\mathrm{v}^{\alpha}_{-\mathbf{q}}\right]_{m'',m'}\left[\mathrm{v}^{\beta}_{0}\right]_{m',m}\left[\mathrm{v}^{\gamma}_{\mathbf{q}}\right]_{m,m''}\nonumber\\
&& \times\frac{1}{E_{m}-E_{m''}}\left[\frac{f(E_{m})-f(E_{m'})}{i\omega_{n}+E_{m}-E_{m'}}+\frac{f(E_{m'})-f(E_{m''})}{i\omega_{n}+E_{m''}-E_{m'}}\right]\nonumber\\
\nonumber\\
\end{eqnarray} 
\begin{eqnarray}
&&T\mathrm{Tr}\left[G\mathrm{v}^{\alpha}_{-q}G\mathrm{v}^{\gamma}_{\mathbf{q}}G\mathrm{v}^{\beta}_{i\omega_{n}}\right]=\nonumber\\
&&\ \ \ \ \ \ \ \ \  \sum_{m,m',m''}\left[\mathrm{v}^{\alpha}_{-\mathbf{q}}\right]_{m'',m'}\left[\mathrm{v}^{\beta}_{0}\right]_{m,m''}\left[\mathrm{v}^{\gamma}_{\mathbf{q}}\right]_{m',m}\nonumber\\
&& \times\frac{1}{E_{m}-E_{m'}}\left[\frac{f(E_{m''})-f(E_{m})}{i\omega_{n}+E_{m''}-E_{m}}+\frac{f(E_{m'})-f(E_{m''})}{i\omega_{n}+E_{m''}-E_{m'}}\right]\nonumber\\
\nonumber\\
\end{eqnarray} 
in which
\begin{eqnarray}
\left[\mathrm{v}^{\alpha}_{\mathbf{q}}\right]_{m,m'}&=&\sum_{\mathbf{p},\sigma} \psi^{m*}_{\mathbf{p+q\sigma}}\left[\mathrm{v}_{\mathbf{q}}^{\alpha}\right]_{\mathbf{p+q,p}}\psi^{m'}_{\mathbf{p}\sigma}\nonumber\\
\left[\mathrm{v}^{\alpha\beta}_{\mathbf{q}}\right]_{m,m'}&=&\sum_{\mathbf{p},\sigma}\psi^{m*}_{\mathbf{p+q\sigma}}\left[\mathrm{v}_{\mathbf{q}}^{\alpha\beta}\right]_{\mathbf{p+q,p}}\psi^{m'}_{\mathbf{p}\sigma}
\end{eqnarray}
is the matrix element of the current and inverse effective mass operator in the basis spanned by the single particle eigenstate $\varphi^{m}_{\mathbf{q},\sigma}$. $f(E_{m})$ is the fermi distribution function at temperature $T$.

After the analytic continuation, we find that the optical conductivity is given by
\begin{eqnarray}
\sigma^{xx}(\omega)&=&-\frac{\mathrm{Im} K^{xx}(\omega+i0^{+})}{\omega}\nonumber\\
&=&\langle \  \frac{\pi}{\omega}\sum_{m,m'}|[\mathrm{v}^{x}_{0}]_{m,m'}|^{2}\times[f(E_{m'})-f(E_{m})]\nonumber\\
 &\times&\delta(\omega-[E_{m}-E_{m'}])  \ \rangle
\end{eqnarray}
The analytic continuation of the Hall conductivity is straightforward can be done by the replacement $i\omega_{n}\rightarrow \omega+i0^{+}$. The resultant expression for the real and the imaginary part of $\sigma^{xy}$ is rather lengthy and not very instructive. Here we will omit it.

We note that the formulas for the optical conductivity and the Hall conductivity derived in this subsection reduce to those derived in Ref.[\onlinecite{Metzner}] for a spiral magnetic ordered state if we ignore the ensemble average and set $\bm{\phi}$ to be the order parameter field describing such a spiral magnetic order. For this reason, we will adopt these formulas in the last subsection to calculate the optical conductivity and Hall conductivity of a system with antiferromagnetic order(a special case of the spiral magnetic order assumed in Ref.[\onlinecite{Metzner}]).

\subsection{Effective action for the thermal fluctuation of the local moment and its regulation on finite lattice}
To describe the thermal fluctuation of the local moment, we adopt a spin susceptibility of the following form
\begin{equation}
\chi(\mathbf{q},i\omega_{n})=\frac{\chi_{0}}{\xi^{-2}+(\mathbf{q}-\mathbf{Q})^{2}}\delta_{i\omega_{n},0}
\end{equation}
The same form has also been adopted in Ref.[\onlinecite{Lin}] in their perturbative study of the effect of thermal fluctuation. Susceptibility of this form can be derived from the standard MMP susceptibility by setting $\chi_{Q}=\chi_{0}\xi^{2}$ and taking the limit of $\omega_{sf}/T\rightarrow 0$. 

In real space, the effective action of the fluctuating local moment reads 
\begin{equation}
S_{eff}[\bm{\phi}]=\frac{1}{2T}\sum_{i,j}\chi^{-1}_{i,j}\bm{\phi}_{i}\cdot\bm{\phi}_{j}
\end{equation}
Here $\chi^{-1}_{i,j}$ is given by
\begin{equation}
\chi^{-1}_{i,j}=\frac{1}{N}\sum_{\mathbf{q}}\chi ^{-1}(\mathbf{q})e^{i\mathbf{q}\cdot(\mathbf{r}_{i}-\mathbf{r}_{j})}
\end{equation}
To perform numerical simulation, we put the system on a finite cluster of square lattice with periodic boundary condition imposed in both the $x$ and the $y$-direction. To be compatible with such a lattice regulation, we make the following replacement in the phenomenological susceptibility
\begin{equation}
(\mathbf{q}-\mathbf{Q})^{2}\rightarrow 4+2[\cos(q_{x})+\cos(q_{y})]
\end{equation}
which becomes exact in the $\mathbf{q}\rightarrow\mathbf{Q}$ limit. After such a lattice regulation, we have
\begin{equation}
\chi^{-1}_{i,j}=\frac{1}{\chi_{0}}[(4+\xi^{-2})\delta_{i,j}+\delta_{\mathbf{r}_{i}-\mathbf{r}_{j},\bm{\delta}}]
\end{equation}
in which $\bm{\delta}$ denotes the four nearest-neighboring vectors on the square lattice. The effective action for $\bm{\phi}$ then becomes
\begin{equation}
S_{eff}[\bm{\phi}]=\frac{(4+\xi^{-2})}{2T\chi_{0}}\sum_{i}\bm{\phi}_{i}\cdot\bm{\phi}_{i}+\frac{1}{2T\chi_{0}}\sum_{i,\bm{\delta}}\bm{\phi}_{i}\cdot\bm{\phi}_{i+\bm{\delta}}
\end{equation}

To be complete, we have also computed the EM response of the system when a non-Gaussian action with the following form is assumed for the fluctuating local moment
\begin{equation}
S_{eff}[\bm{\phi}]=\eta \sum_{i,\bm{\delta}}\bm{\phi}_{i}\cdot\bm{\phi}_{i+\bm{\delta}}
\end{equation}
The non-Gaussian nature of this action is manifested in the following constraint on the fluctuating field $\bm{\phi}_{i}$ 
\begin{equation}
\bm{\phi}_{i}\cdot\bm{\phi}_{i}=1
\end{equation}
$\eta$ is a constant introduced to tune the correlation length of the fluctuating moment. This action is called the non-linear sigma model(NLSM) in the literature and it describes the thermal spin fluctuation in the renormalized classical regime of a two dimensional quantum Heisenberg antiferromagnet\cite{CHN}. The NLSM description is thought to be relevant to the electron-doped cuprate superconductors\cite{Damascelli}, in which the spin fluctuation is longer ranged and closer to the static limit.

The distribution as described by the Gaussian action Eq.74 can be sampled with either the standard Metropolis algorithm, or more directly, a direct sampling of the multiple Gaussian distribution. Since the parameter $T\chi_{0}$ in the action can be absorbed by a redefinition of the integration variable $\bm{\phi}$, with the only effect being to replace the parameter $U$ in $\mathcal{H}[\psi,\psi^{\dagger},\bm{\phi},\mathbf{A}=0]$ by $\tilde{U}=U\sqrt{T\chi_{0}}$, we will set $T\chi_{0}=1$ and treat $\tilde{U}$ as a free parameter. $\tilde{U}$ measures the overall strength of the coupling between the antiferromagnetic spin fluctuation and the electron. To sample the non-Gaussian action of the NLSM as give by Eq.75, we adopt the heat bath algorithm\cite{Young} supplemented with the over-relaxation trick on the unit vector $\bm{\phi}_{i}$\cite{Okubo}. The desired correlation length $\xi$ is achieved by tuning the value of the parameter $\eta$ in the action. 

\subsection{Numerical results}
Our numerical calculation is performed on a finite cluster of square lattice with $L\times L=400$ sites. Here $L$ is the linear size of the cluster. Periodic boundary condition will be assumed in both the $x$ and the $y$-direction of the cluster. We will consider a tight binding model with only nearest neighboring hopping $t$ and next-nearest neighboring hopping $t'$. We set $t'=-0.3t$ and $\mu=-t$, which corresponds to a hole doping level of about $15\%$ in the non-interacting limit. In the following, we will use $t$ as the unit of energy. The temperature is set to be $T=0.1t$ in the calculation. We have generated 6400 statistically independent samples from the distribution $e^{-S_{eff}[\bm{\phi}]}$ for the fluctuating local moment $\bm{\phi}$. The optical conductivity and the Hall conductivity are computed by averaging the corresponding fermion traces appearing in Eq.59 and Eq.62 over these samples. We should keep in mind that to fix the doping level in the calculation we must adjust the chemical potential with both the temperature and the coupling strength $\tilde{U}$. Such a more detailed treatment is not attempted here, since our goal is restricted to a qualitative discussion of the effect of the thermal spin fluctuation, rather than a quantitative fitting of experimental data.

\subsubsection{Optical conductivity}

The calculated optical conductivity when we assume a Gaussian action of the form Eq.74 is shown in Fig.1. Here we treat $\tilde{U}$ and $\xi$ as independent parameters, although in reality both of them may have involved temperature and doping dependence. For small $\tilde{U}$, $\sigma^{xx}(\omega)$ is seen to be dominated by a Drude peak at low energy. This is to be contrasted with the behavior at large $\tilde{U}$, in which case the optical spectrum is dominated by a broad mid-infrared peak that extends to the energy scale of the band width. In between these two limits, the optical spectrum exhibits a two-component structure with a Drude component at low frequency and a mid-infrared component at higher frequency. Such a two-component character becomes increasingly more evident with the increase of the correlation length of the local moment. The Drude component can be attributed to the residual electron density of state near the fermi level in the presence of the thermal spin fluctuation. More specifically, it is derived from the intra-band optical weight when the fluctuating local moment condenses into a static antiferromagnetic long range order and that the electron band split into well defined upper and lower SDW band. Correspondingly, the mid-infrared component is derived from the inter-band optical weight between such SDW split bands. A more detailed discussion on this point is presented in the last subsection of this section.       

\begin{figure}
\includegraphics[width=6.5cm]{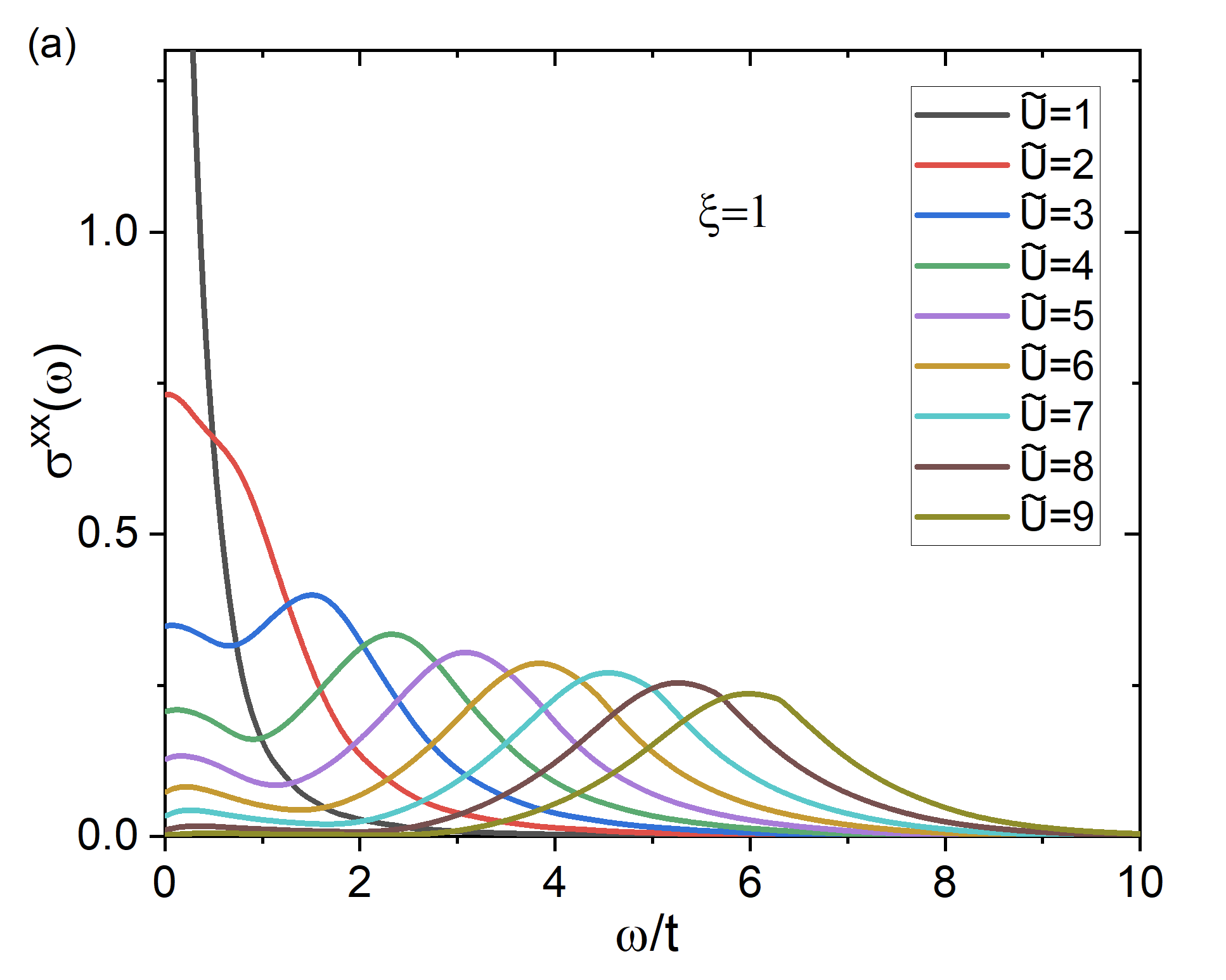}
\includegraphics[width=6.5cm]{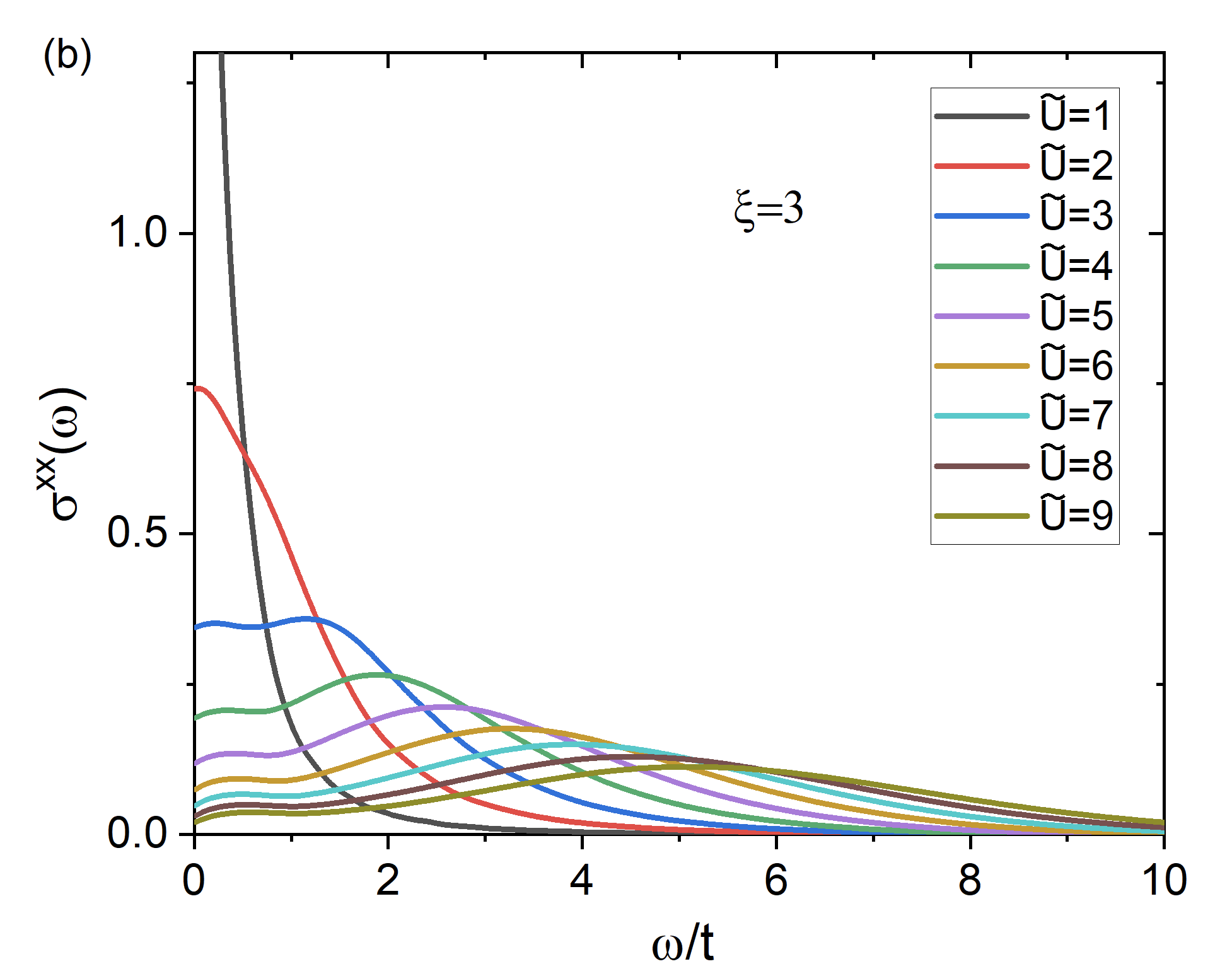}
\includegraphics[width=6.5cm]{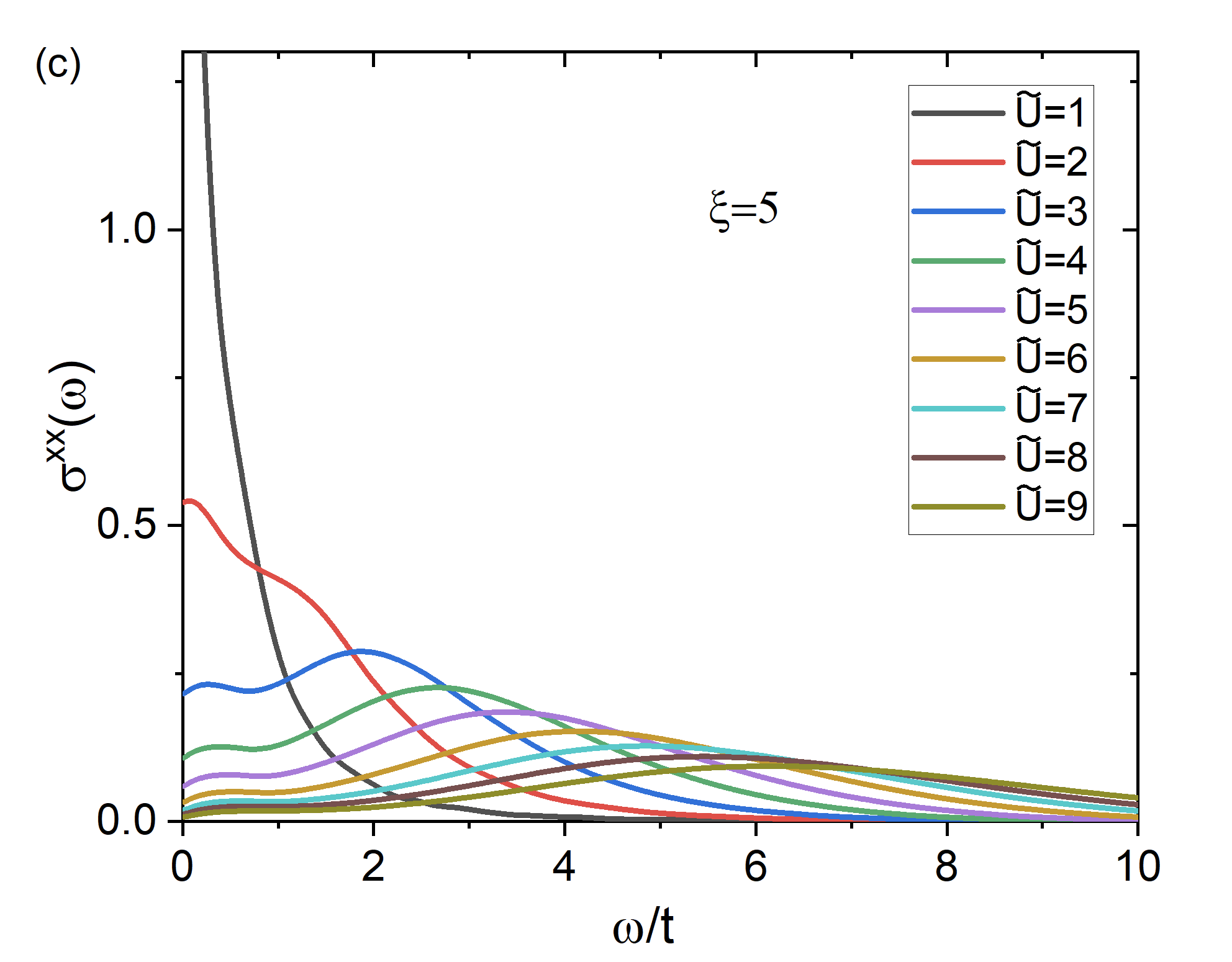}
\includegraphics[width=6.5cm]{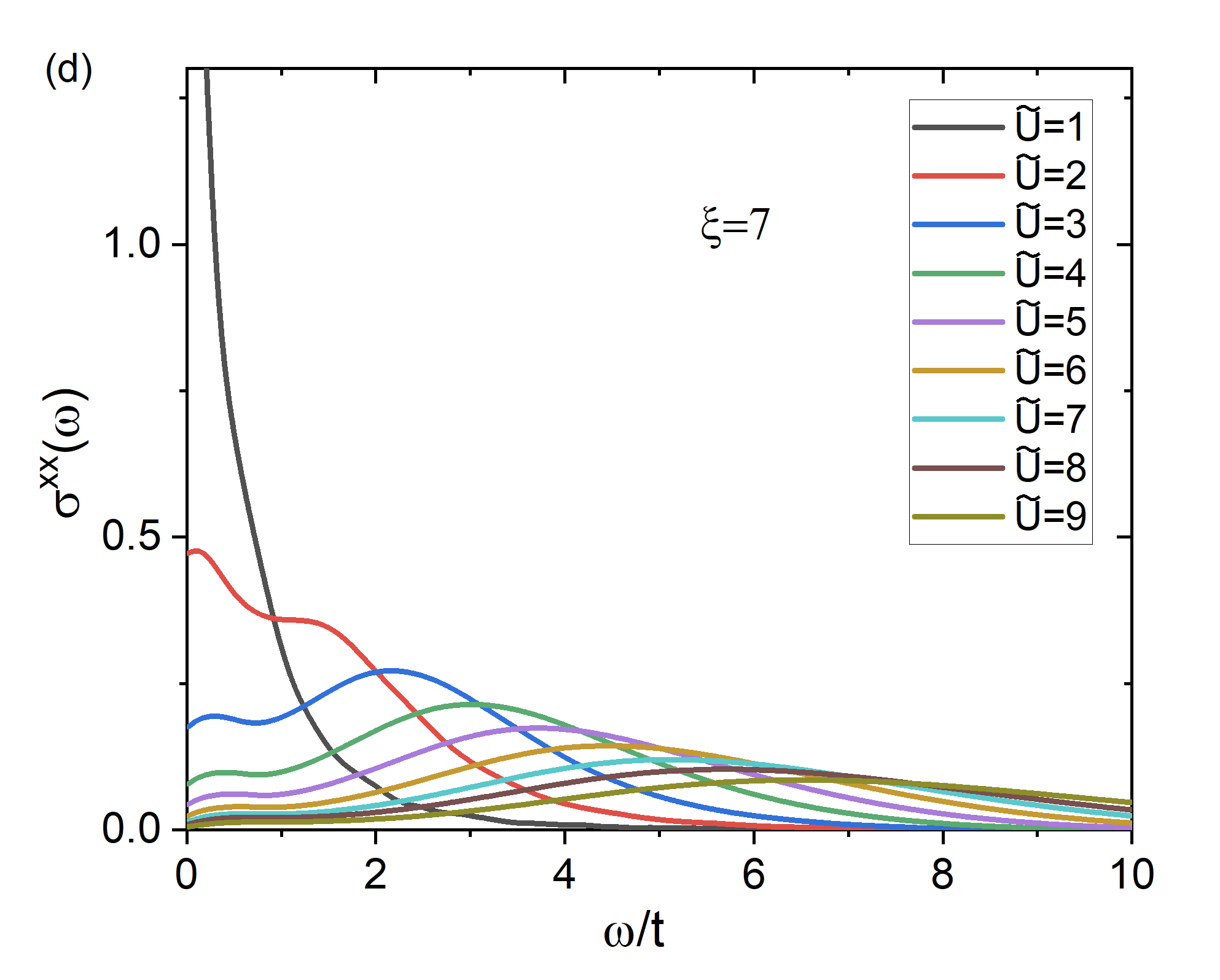}
\caption{The optical conductivity of the system when we assume a Gaussian action of the form Eq.74 for the thermally fluctuating local moment. Here we treat $\tilde{U}$ and $\xi$ as independent parameters, although in reality both of them may have involved temperature and doping dependence. The energy is measured in unit of $t$ and we have set $k_{B}T=0.1t$ for the electron. The $\delta$-function peak is broadened into a Lorentzian peak of width $0.03t$ in the calculation. The calculation is done on a finite cluster of square lattice with $L\times L=400$ sites and periodic boundary condition in both the $x$ and the $y$-direction.}
\end{figure}  

A two-component structure has been indeed ubiquitously observed in the optical spectrum of the cuprate superconductors\cite{Basov,Marel,Heumen}. However, the Drude peak predicted here is somehow too weak to be consistent with the experimental observations in the strong coupling regime. Such a discrepancy should be attributed to the neglect of the quantum nature of the local moment fluctuation. In fact, the neglect of $\tau$-dependence in $\bm{\phi}_{i}(\tau)$ becomes invalid for electron transition below the characteristic frequency of the local moment fluctuation. The quantum nature of the spin fluctuation becomes important in such a situation. This is particularly important in the hole-doped cuprate superconductors in which the spin fluctuation is more dynamic than that in the electron-doped cuprate superconductors. Such a quantum effect is expected to recover partially the electron density of state near the fermi level from the SDW gapping by the thermal spin fluctuation and is thus expected to enhance the spectral weight contained in the Drude peak. On the other hand, the mid-infrared spectral weight is not expected to be significantly influenced by such an effect since it is located at a frequency significantly higher than that of the local moment fluctuation. 
 
As a comparison, we present the optical conductivity calculated when a non-Gaussian action with the form of the NLSM is assumed in Fig.2. When compared to the results we got for the Gaussian action, the two-component character in the optical spectrum becomes even more evident. It is interesting to compare these results with the observations in the electron-doped cuprate superconductors, for which the NLSM description is more appropriate.

\begin{figure}
\includegraphics[width=6.5cm]{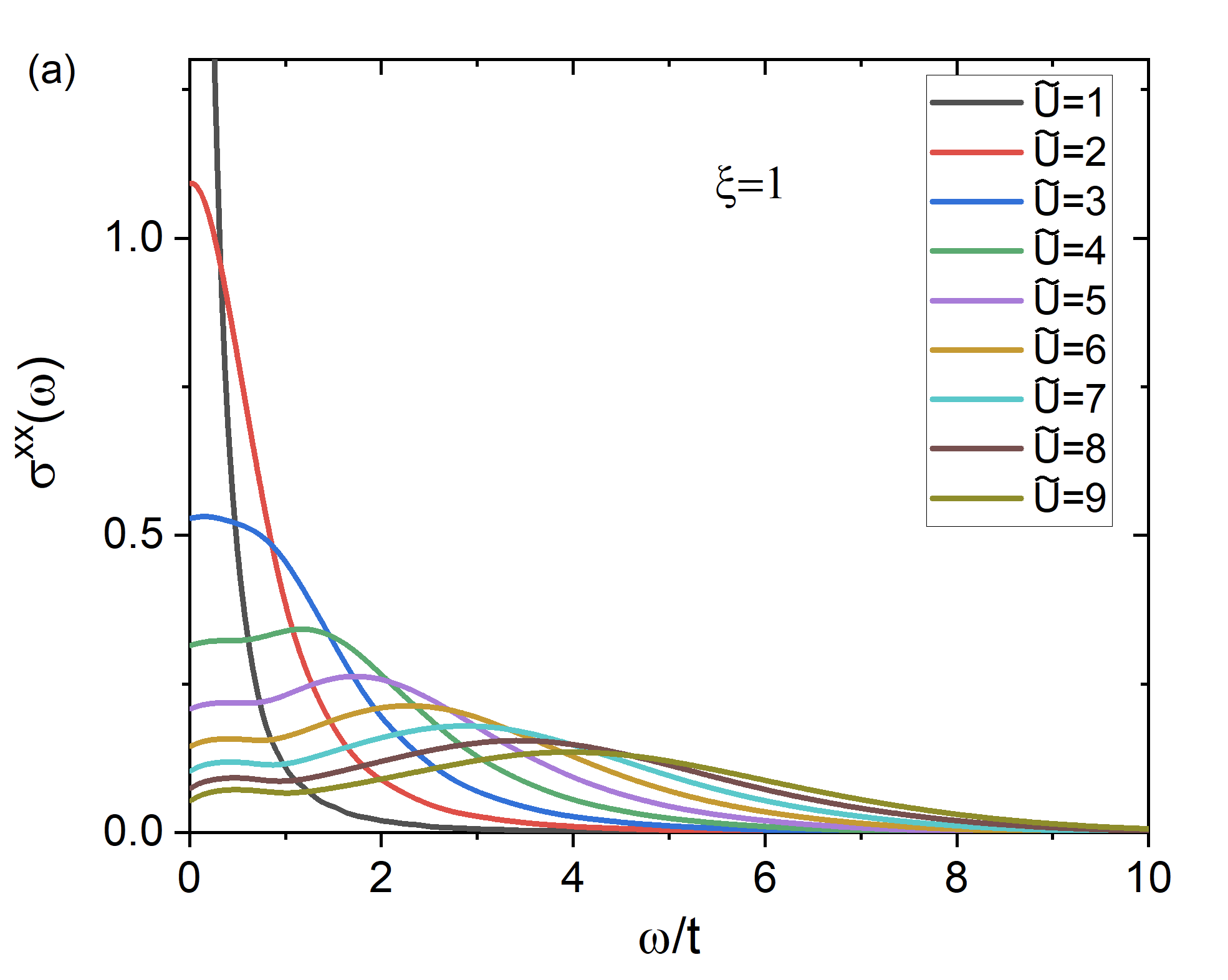}
\includegraphics[width=6.5cm]{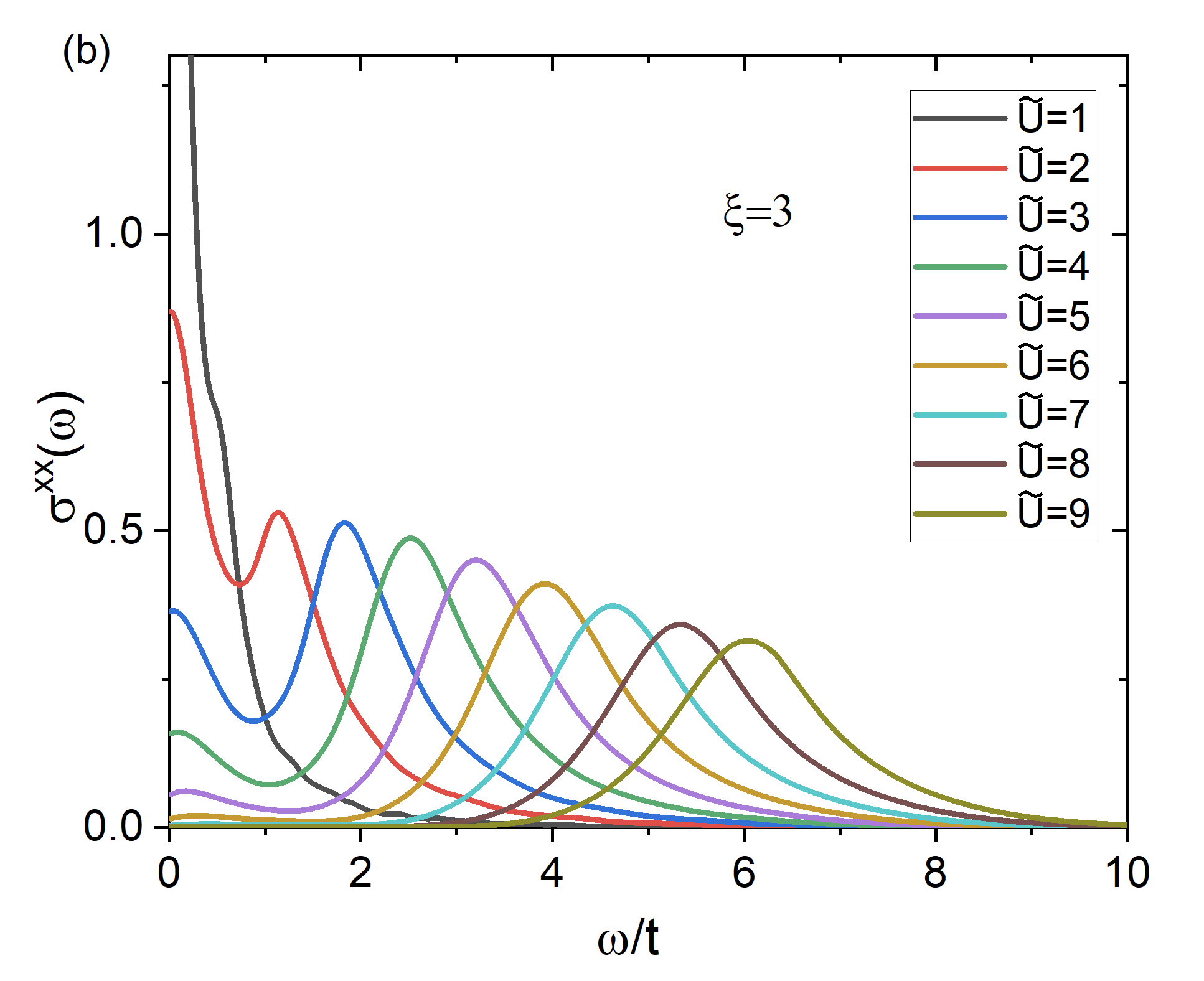}
\includegraphics[width=6.5cm]{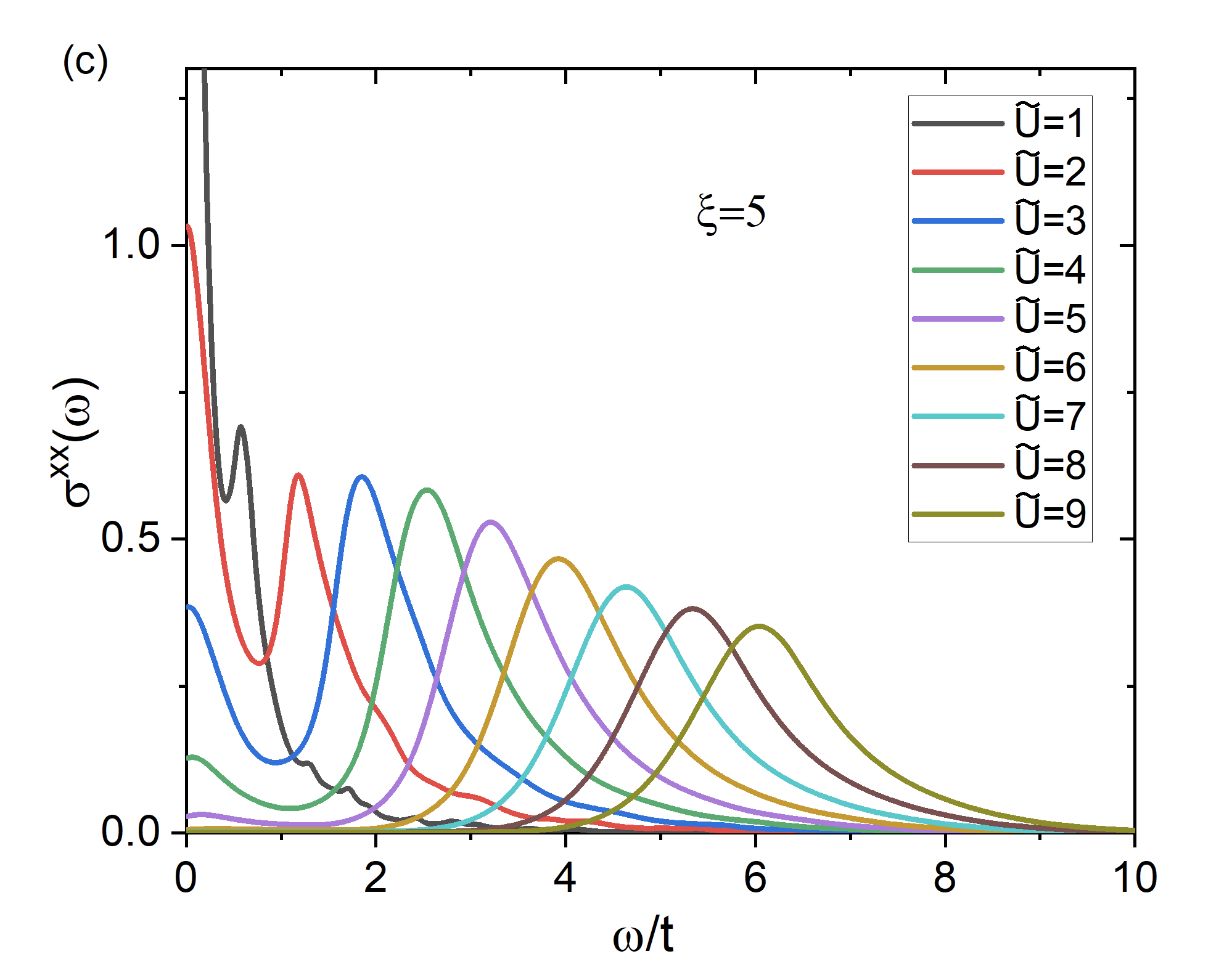}
\includegraphics[width=6.5cm]{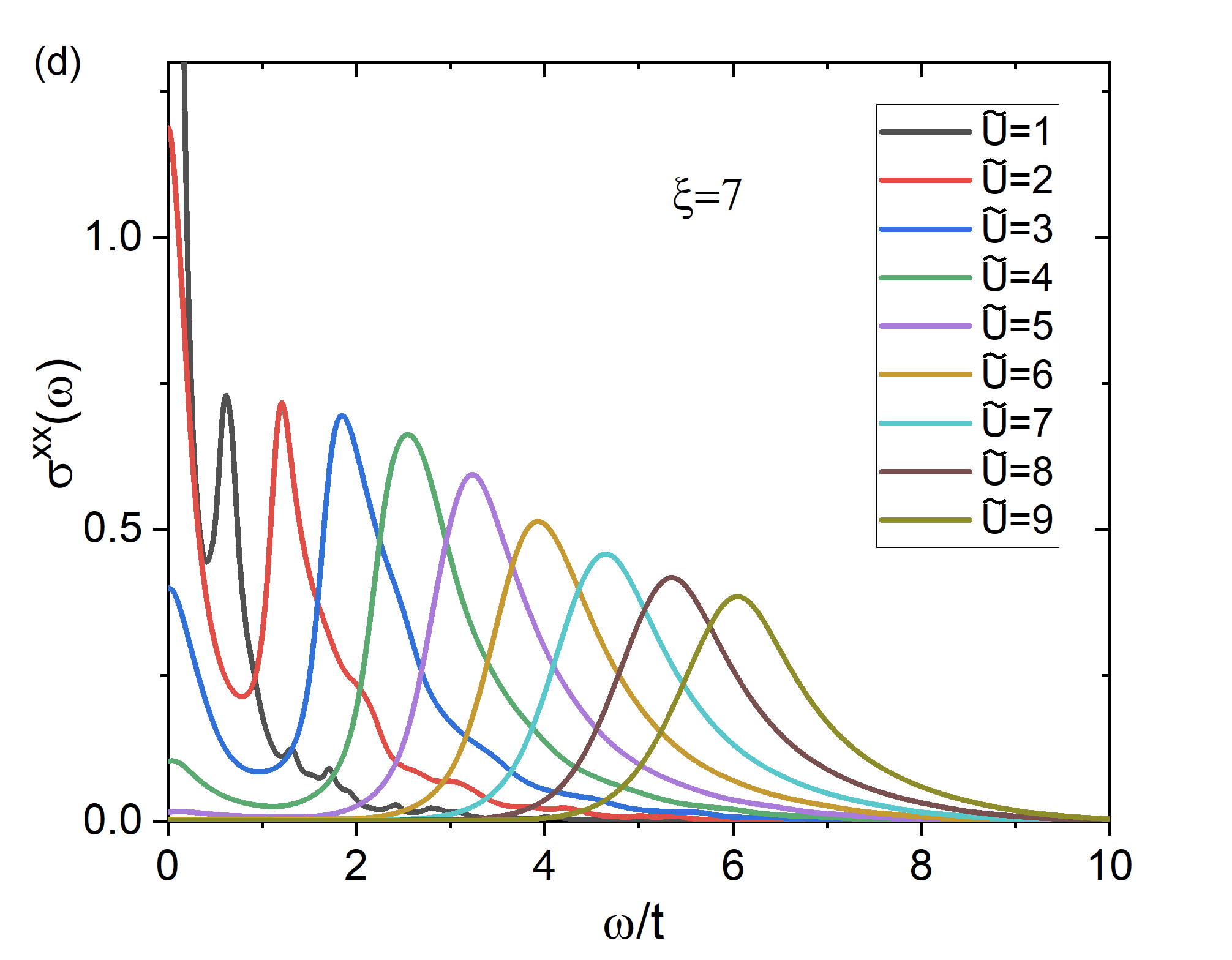}
\caption{The optical conductivity of the system when we assume a non-Gaussian action of the form Eq.75 for the thermally fluctuating local moment. Such an action describes the thermal fluctuation in the renormalized classical regime of a quantum Heisenberg antiferromagnetic model defined on the square lattice. It is thought to be relevant to the electron-doped cuprate superconductors. The energy is measured in unit of $t$ and we have set $k_{B}T=0.1t$ for the electron. The $\delta$-function peak is broadened into a Lorentzian peak of width $0.03t$ in the calculation. The calculation is done on a finite cluster of square lattice with $L\times L=400$ sites and periodic boundary condition in both the $x$ and the $y$-direction.}
\end{figure}  

We note that the optical conductivity calculated here is free from the unphysical divergence in the low frequency regime encountered in the perturbative treatment\cite{Lin}. This is crucial to reveal the two-component structure of the optical spectrum. Such a divergence can be traced back to the coalescence of Green's function poles in the vertex correction for the current operator in the long wave length limit. It can be cured only when we take into account contribution at higher perturbative orders in a consistent way, which is a formidable task. Interestingly, we find in a recent work that such a divergence also plays an important role in the vertex correction to the electron-phonon coupling in an antiferromagnetic correlated background\cite{Liu}.  

\subsubsection{Hall conductivity}
While the Hall response in the DC limit is more extensively studied in the experimental literature, here we will focus on the Hall response in the optical frequency regime. The rationale for such a choice is twofold. First, while the DC Hall response is more accessible experimentally, its theoretical computation and analysis is more involved. In fact, the DC Hall response of an electron system depends sensitively on details of the low energy of the system. For example, in the Boltzmann transport regime for well defined quasiparticles, it is well known that the DC Hall conductivity of the system depends sensitively on the geometry of the fermi surface and the momentum dependence of the quasiparticle life time on the fermi surface\cite{Ong}. The situation become even more subtle for a strongly correlated system like the cuprate superconductor in which no well-defined quasiparticle exist\cite{Singh,Imada,Shastry,Auerbach,Huang,Wang}. This point is well demonstrated by the calculation below, from which we will see that not only the magnitude but also the sign of the DC Hall conductivity can change dramatically with the system parameters. Second, as we emphasized in the last subsection, the assumption of the dominance of the thermal fluctuation of the local moment will eventually break down when we consider the EM response of the system in the low frequency limit. Thus while the DC Hall response calculated from our framework is free from the unphysical divergence encountered in the perturbative treatment, its relevance to the real cuprate material is questionable.      

Unlike the DC Hall response, the Hall response in the optical frequency regime exhibits more universal trends that is closely related to the strong correlation effect\cite{infrared2,RSI,infrared3,infrared4,infrared5,Lin}. Previously, the Hall response in the optical frequency regime has been measured in various cuprate superconductors\cite{infrared0,infrared1,infrared2,RSI,infrared3,infrared4,infrared5}. However, due to the difficulty in the analysis of these data, the full implication of these results has not been thoroughly investigated. In the measurement of the optical Hall response, the key quantity is the imaginary part of the Hall conductivity $\mathrm{Im}\sigma^{xy}(\omega)$. We will call it the Hall conductivity spectrum and will focus on it below.

The Hall conductivity is calculated from Eq.61 to Eq.66 numerically on a finite cluster of square lattice of linear size $L=20$ with periodic boundary condition. The derivative of $\tilde{\kappa}^{\alpha\beta\gamma}(q_{x},\omega+i0^{+})$ with respect to $q_{x}$ in Eq.61 is approximated by the following finite difference 
\begin{eqnarray}
&&\frac{\partial }{\partial q_{x}}\tilde{\kappa}^{\alpha\beta\gamma}(q_{x},\omega+i0^{+}) |_{q_{x}\rightarrow0}\nonumber\\
&&\ \ \ \ \ \approx \frac{1}{q_{m}}\left[ \tilde{\kappa}^{\alpha\beta\gamma}(q_{x}=q_{m},\omega+i0^{+})-\tilde{\kappa}^{\alpha\beta\gamma}(0,\omega+i0^{+}))\right]\nonumber\\
\end{eqnarray}
in which $q_{m}=2\pi/L$ is the smallest nonzero momentum in the Brillouin zone for the $L\times L$ cluster. Gauge invariance dictates that $\tilde{\kappa}^{\alpha\beta\gamma}(0,\omega+i0^{+}))\equiv 0$. We have checked numerically that this is indeed the case. Thus we have 
\begin{eqnarray}
&&\frac{\partial }{\partial q_{x}}\tilde{\kappa}^{\alpha\beta\gamma}(q_{x},\omega+i0^{+}) |_{q_{x}\rightarrow0}\nonumber\\
&&\ \ \ \ \ \ \ \ \ \ \approx \frac{1}{q_{m}} \tilde{\kappa}^{\alpha\beta\gamma}(q_{x}=q_{m},\omega+i0^{+})\nonumber\\
\end{eqnarray}
However, the introduction of the small but nonzero wave vector $q_{x}=q_{m}$ will break the rotational symmetry about the $z$-axis. Thus the Hall conductivity $\sigma^{\alpha\beta}$ calculated from the above finite difference approximation will no longer be strictly antisymmetric with respect to the exchange of $\alpha$ and $\beta$. To solve this problem, we antisymmetrize the calculated Hall conductivity by hand.
 
The Hall conductivity spectrum $\mathrm{Im}\sigma^{xy}(\omega)$ calculated with the Gaussian effective action Eq.74 is shown in Fig.3. Here we use the same set of parameters as we used in the calculation of the optical conductivity. As we find in the optical conductivity, a two-component structure becomes increasingly evident in $\mathrm{Im}\sigma^{xy}(\omega)$ with the increase of the interaction strength $\tilde{U}$ and the the spin correlation length $\xi$. More specifically, $\mathrm{Im}\sigma^{xy}(\omega)$ is seen to be dominated by a low energy peak for small $\tilde{U}$. This is to be contrasted with the behavior at large $\tilde{U}$, in which case $\mathrm{Im}\sigma^{xy}(\omega)$ is dominated by a broad peak starting from the mid-infrared regime and extending to the energy scale of the band width. As will be clear in the next subsection, this low energy peak is derived from the corresponding spectral feature related to intra-band transition in a SDW ordered state. On the other hand, the broad peak starting from the mid-infrared regime is derived from the corresponding spectral feature related to the inter-band transition in such a state.

Unlike the optical conductivity, the Hall conductivity spectrum $\mathrm{Im}\sigma^{xy}(\omega)$ can be either positive or negative. From Fig.3 we see that while the low energy peak and the mid-infrared peak are both positive, there exists two sign changes in $\mathrm{Im}\sigma^{xy}(\omega)$ between these two spectral components when $\tilde{U}$ is large. As we will see in the next subsection, such sign changes are the precursor of the antiferromagnetic long-range ordering and the related fermi surface reconstruction. In fact, the large positive peak at low energy is naturally expected for our system when $\tilde{U}$ is small, since the system in the non-interacting limit possess a large hole fermi surface around $(\pi,\pi)$ as illustrated by the thick blue lines in Fig.6. When we turn on an antiferromagnetic long range order, the large hole fermi surface will be reconstructed with the emergence of four hole pockets around $(\pm\pi/2,\pm\pi/2)$ and two electron pockets around $(0,\pi)$ and $(\pi,0)$(as illustrated by the thick black lines in Fig.6). The two sign changes occurring between the low energy peak and the mid-infrared peak are just the consequence of such fermi surface reconstruction. For our choice of parameters, the hole pocket always dominate the electron pocket in the reconstructed fermi surface. We thus expect the low energy peak in $\mathrm{Im}\sigma^{xy}(\omega)$ to be always positive. This is different from the situation in Ref.[\onlinecite{Lin}], in which the electron pocket dominates the hole pocket. Consequently, the low energy peak in $\mathrm{Im}\sigma^{xy}(\omega)$ can be made negative when $\tilde{U}$ is sufficiently large. We also note that $\mathrm{Im}\sigma^{xy}(\omega)$ approaches zero from the positive side in the high frequency limit.

\begin{figure}
\includegraphics[width=6.5cm]{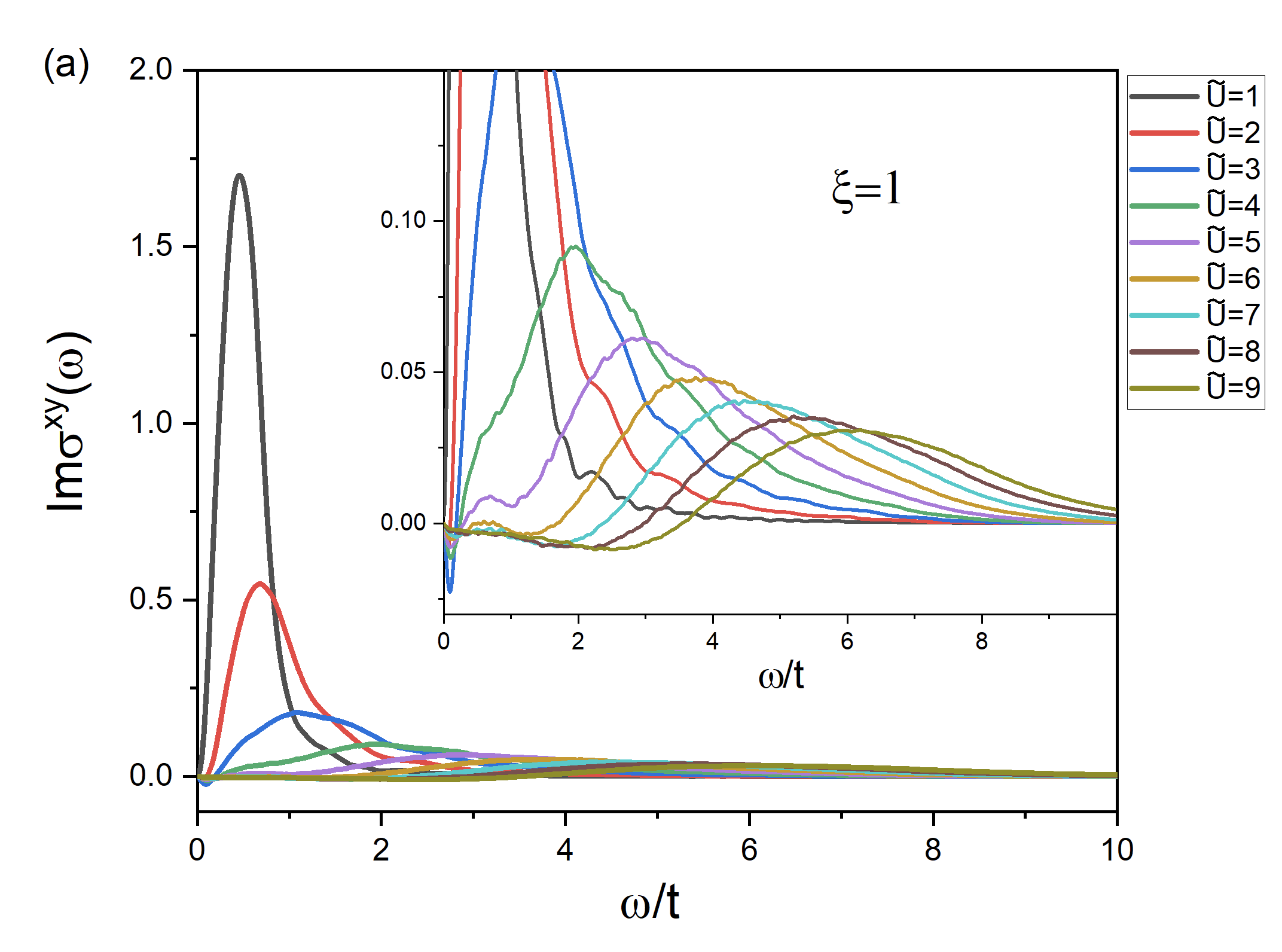}
\includegraphics[width=6.5cm]{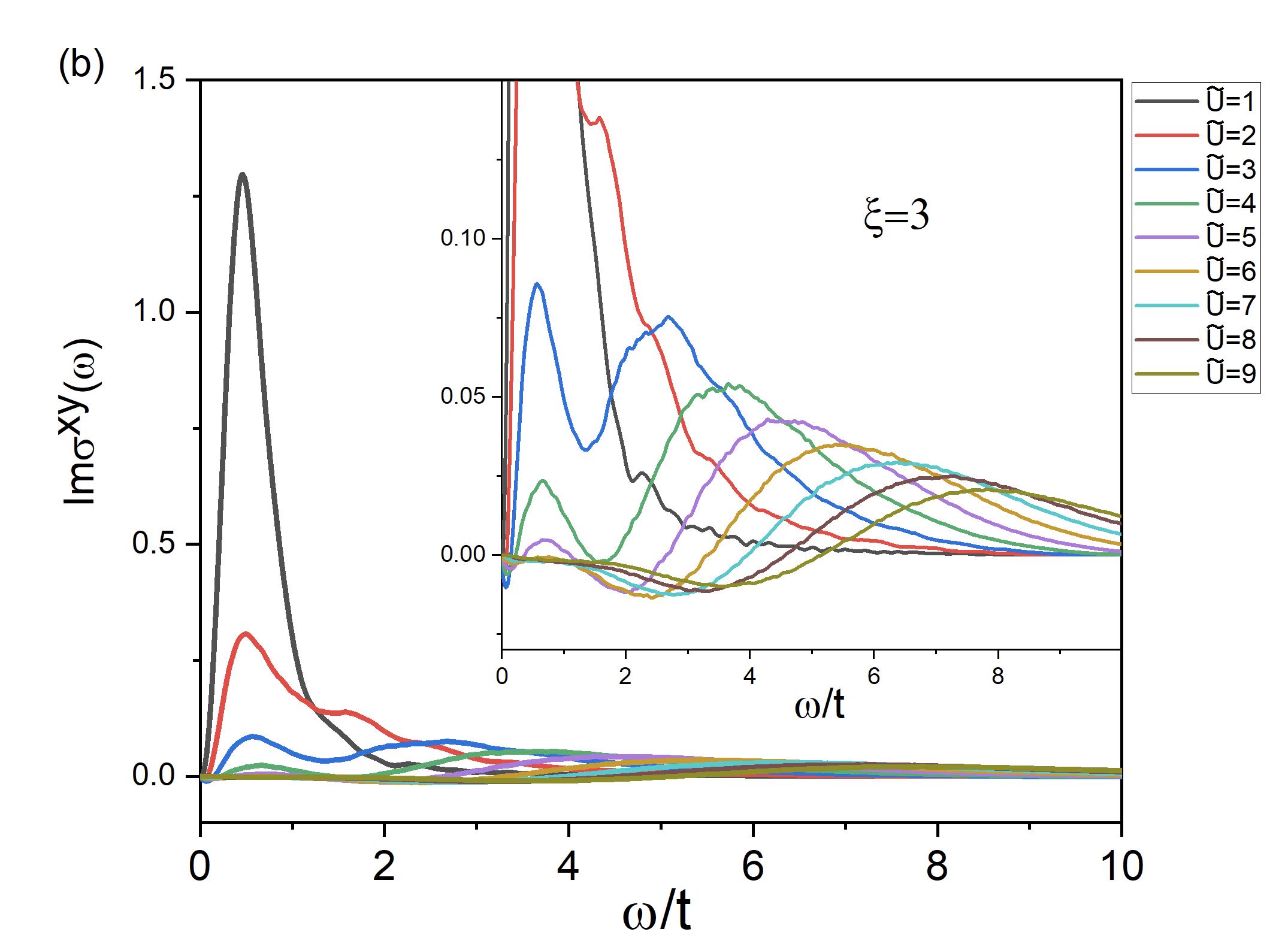}
\includegraphics[width=6.5cm]{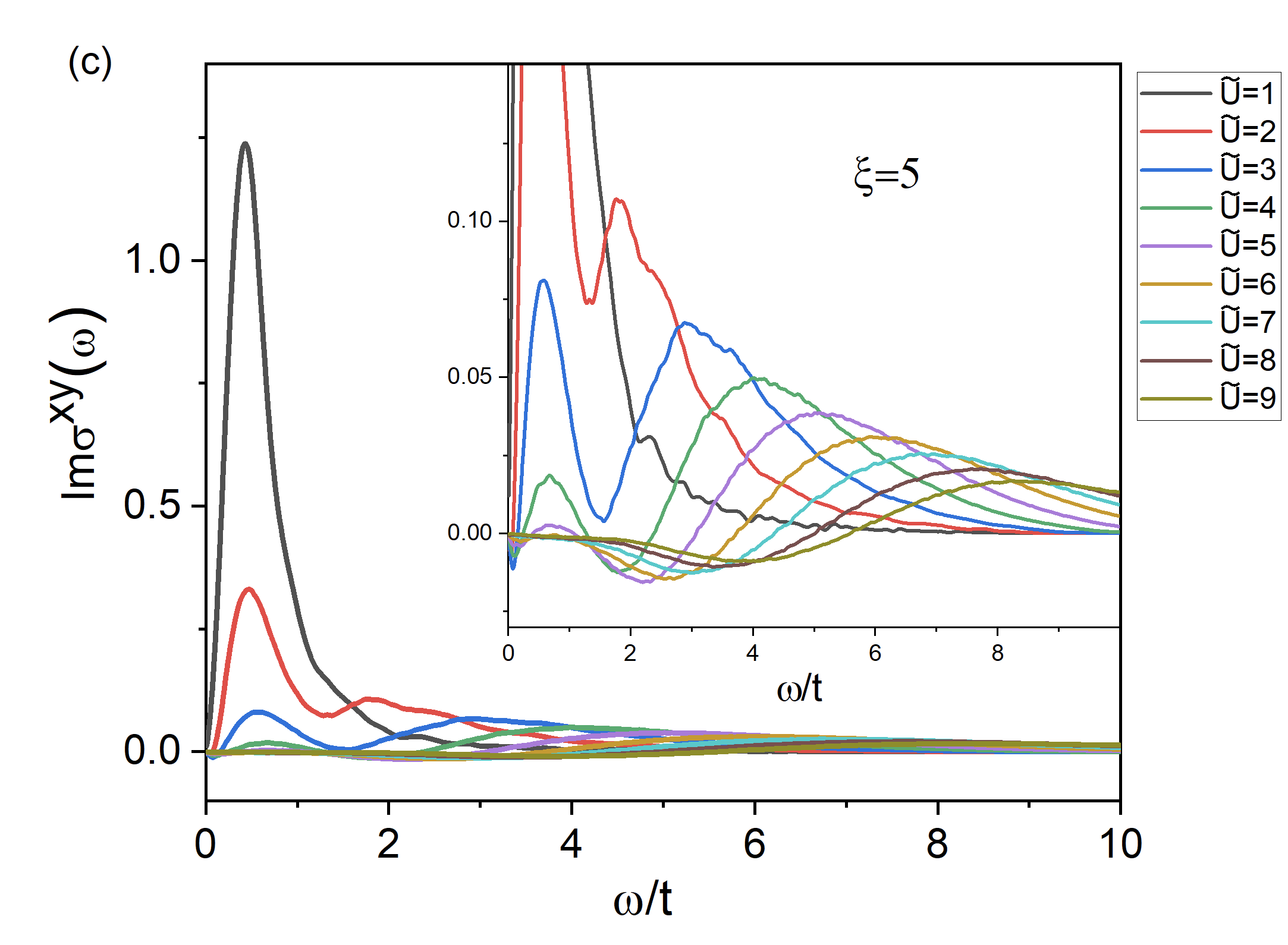}
\includegraphics[width=6.5cm]{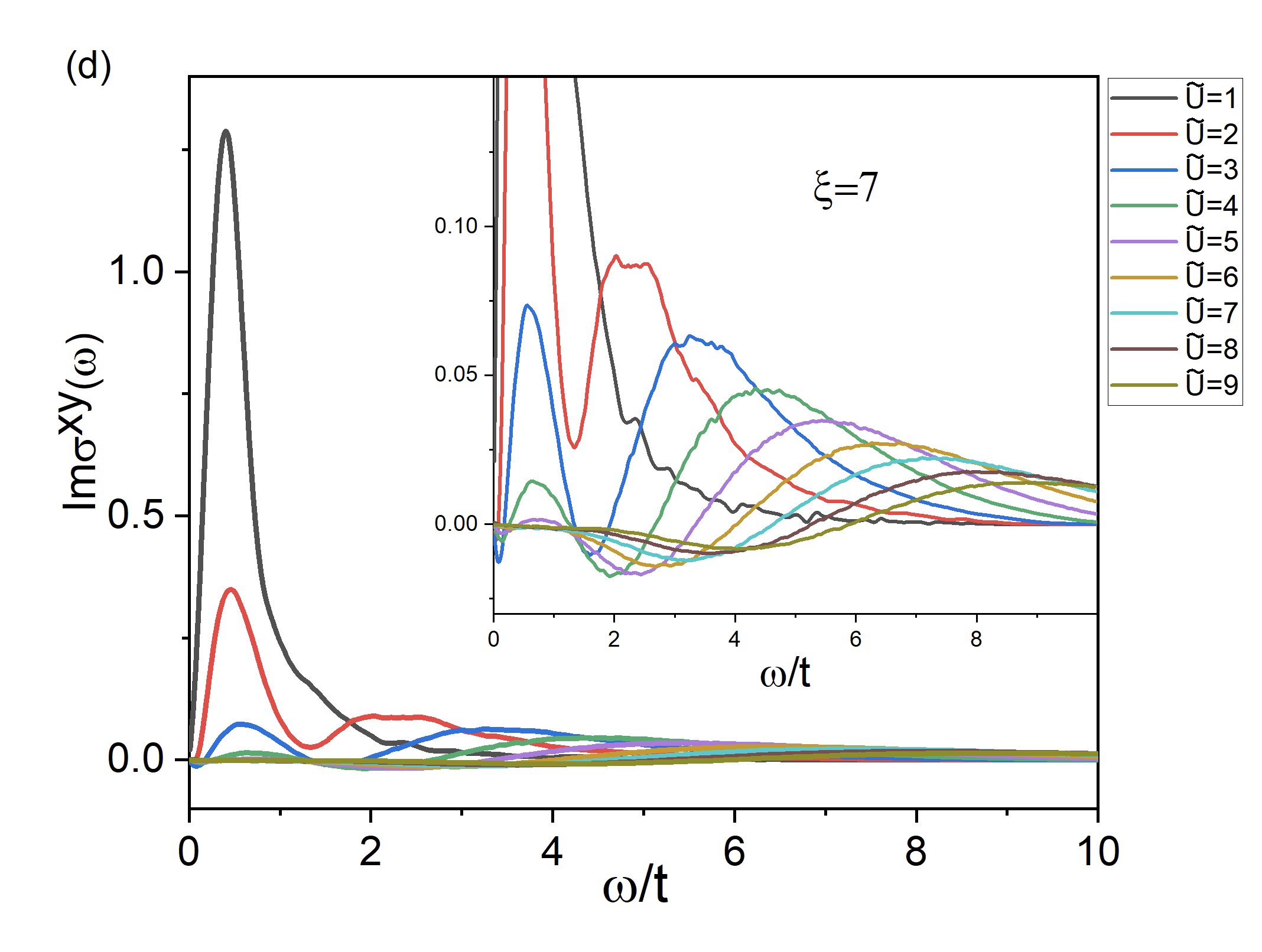}
\caption{The imaginary part of the infrared Hall conductivity calculated with a Gaussian action of the form Eq.74 for the thermally fluctuating local moment. Here we have treated $\tilde{U}$ and $\xi$ as independent parameters, although both of them have involved temperature and doping dependence. The energy is measured in unit of $t$ and and we have set $k_{B}T=0.1t$ for the electron. The $\delta$-function peak is broadened into a Lorentzian peak of width $0.03t$ in the calculation. The calculation is done on a finite cluster of square lattice with $L\times L=400$ sites and periodic boundary condition in both the $x$ and the $y$-direction. The inset represents a zoom in view of the Hall conductivity spectrum. We note that the tiny negative overshoot near $\omega=0$ can be attributed to finite size effect in the anti-symmetrization procedure.}
\end{figure}  

The Hall conductivity spectrum calculated with an effective action of the form of the NLSM is shown in Fig.4. The basic trends are the same as those we find for the Gaussian action. Consistent with our observation on the optical conductivity, the two-component structure in $\mathrm{Im}\sigma^{xy}(\omega)$ is more clearly seen when we assume the NLSM action. To be complete, we also present the real part of the Hall conductivity in Fig.5 for NLSM action and with $\xi=3$. $\mathrm{Re}\sigma^{xy}(\omega)$ for other settings are qualitatively similar. As we find in $\mathrm{Im}\sigma^{xy}(\omega)$, a two-component structure can be clearly identified in $\mathrm{Re}\sigma^{xy}(\omega)$. For large $\tilde{U}$, we find that there are three sign reversals in $\mathrm{Re}\sigma^{xy}(\omega)$. The first two sign reversals occur between the low energy peak and the mid-infrared peak, the third sign reversal occurs above the mid-infrared peak. $\mathrm{Re}\sigma^{xy}(\omega)$ approaches zero from the negative side in the high frequency limit.

 \begin{figure}
\includegraphics[width=6.5cm]{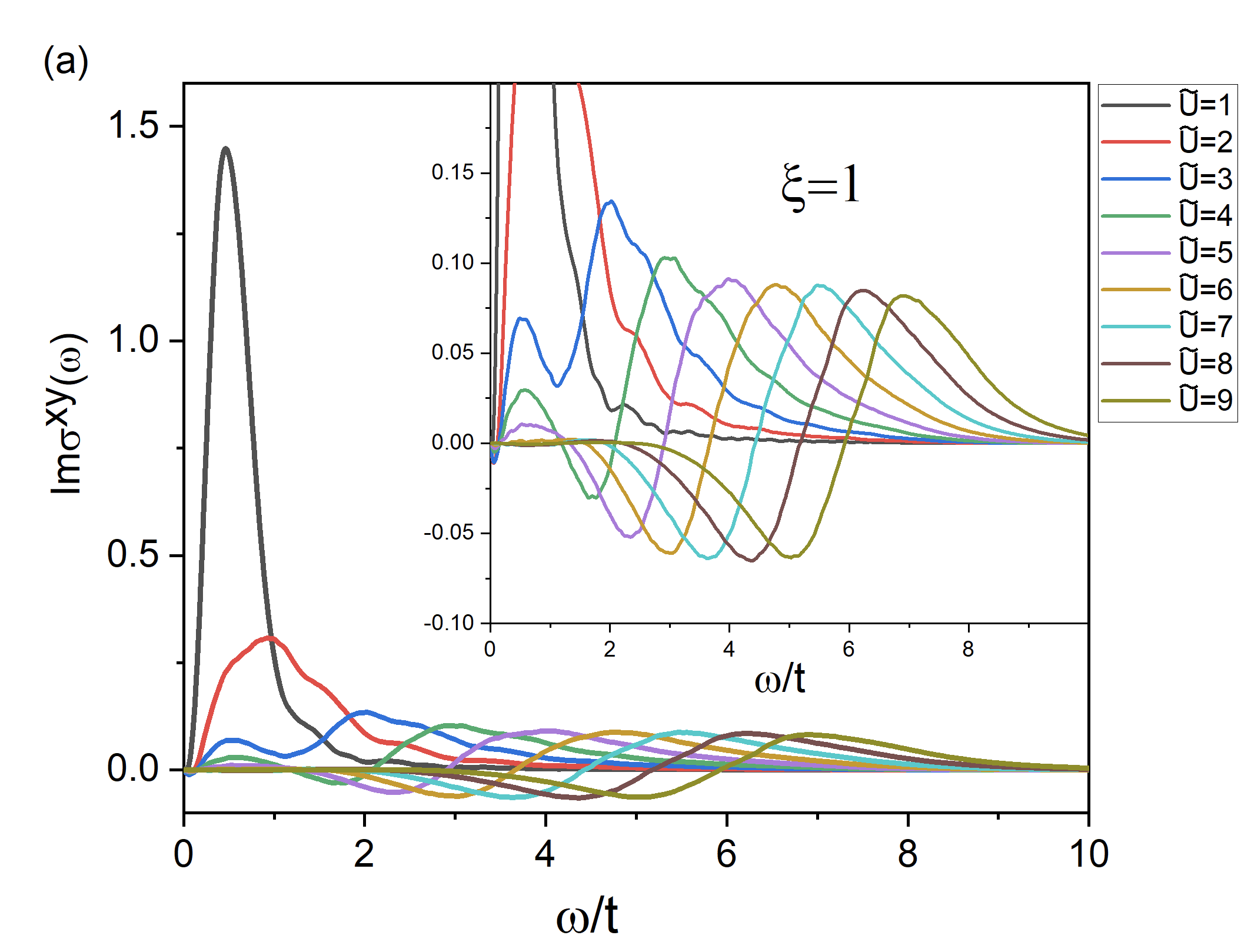}
\includegraphics[width=6.5cm]{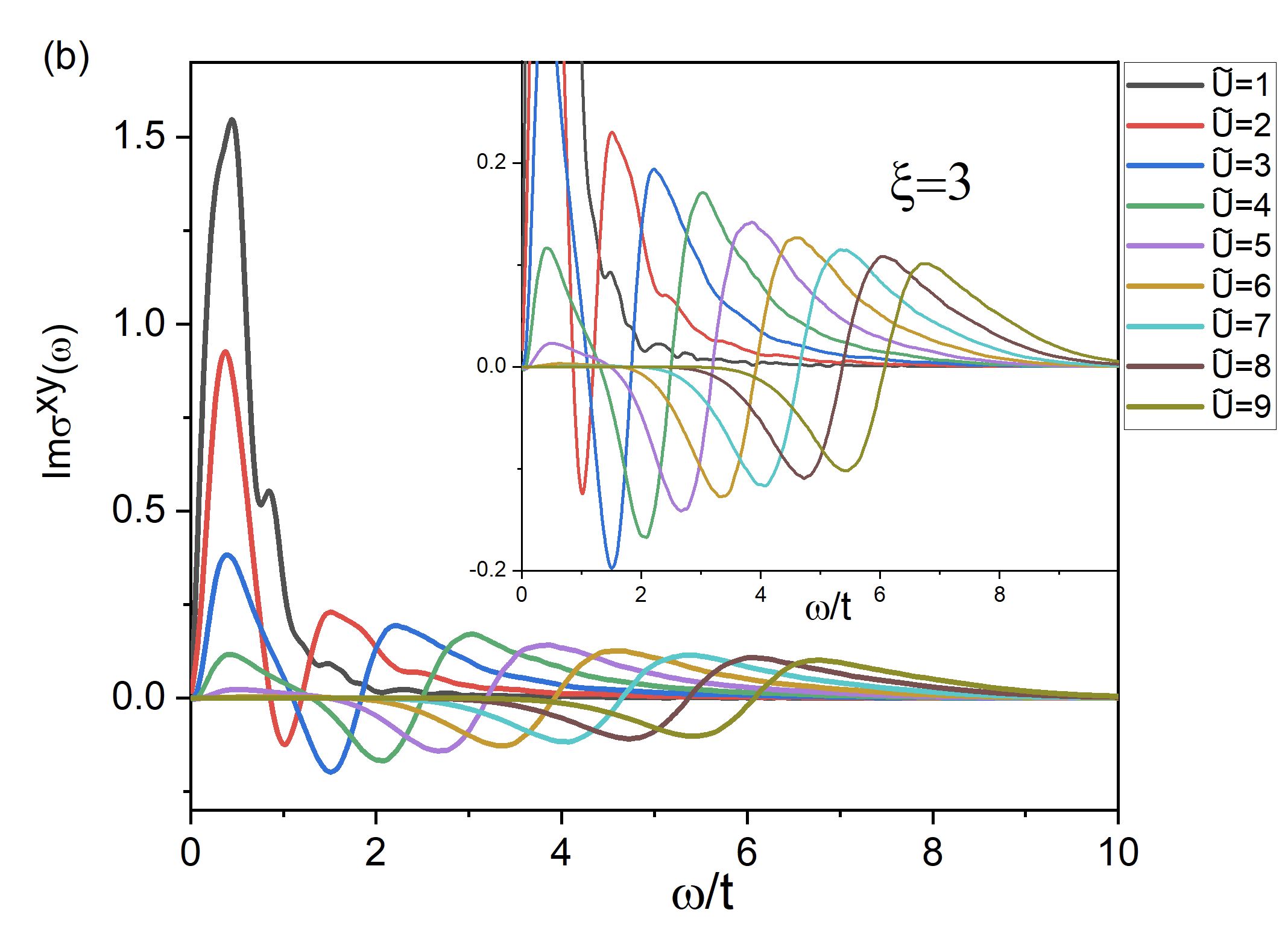}
\includegraphics[width=6.5cm]{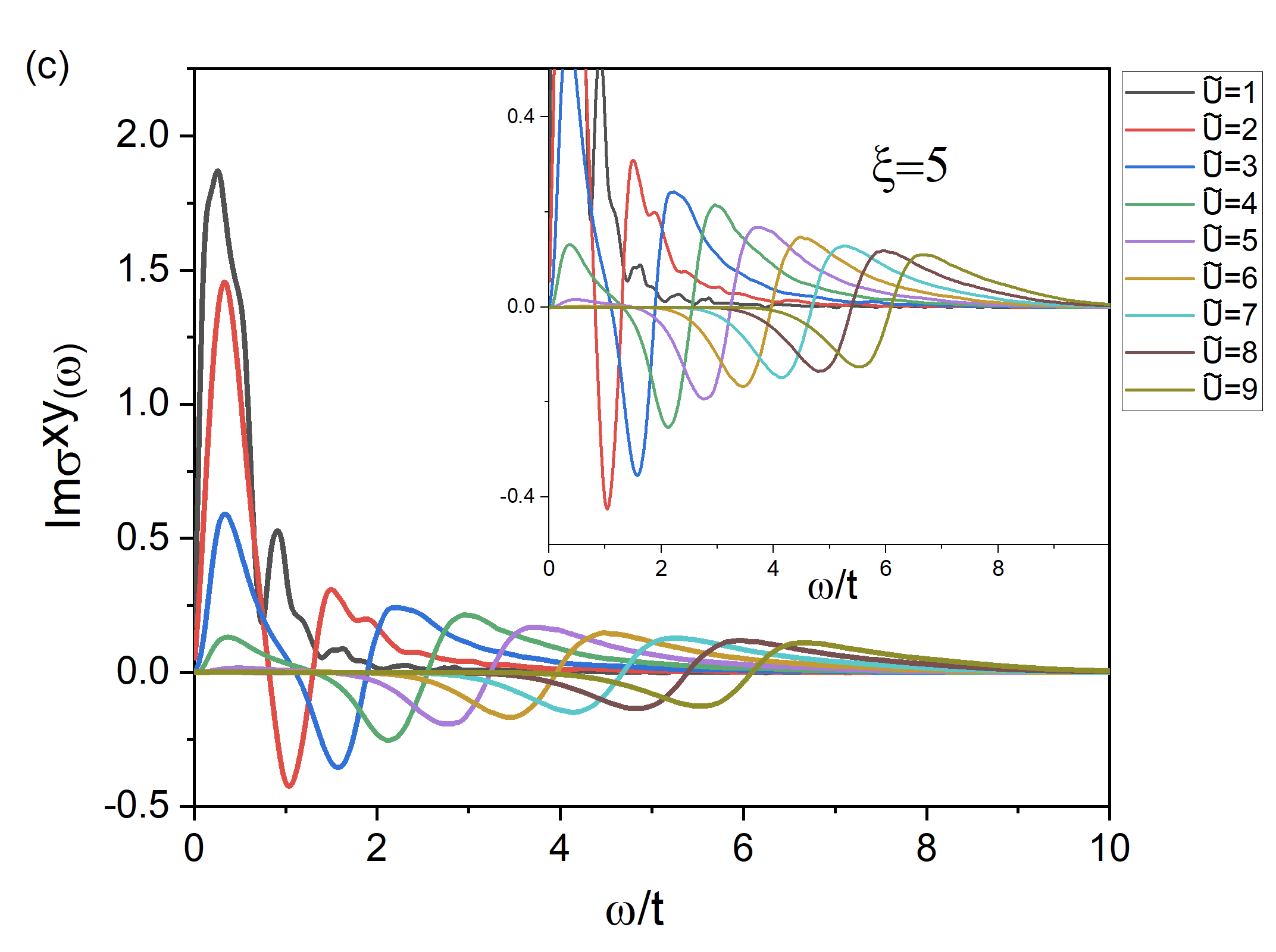}
\includegraphics[width=6.5cm]{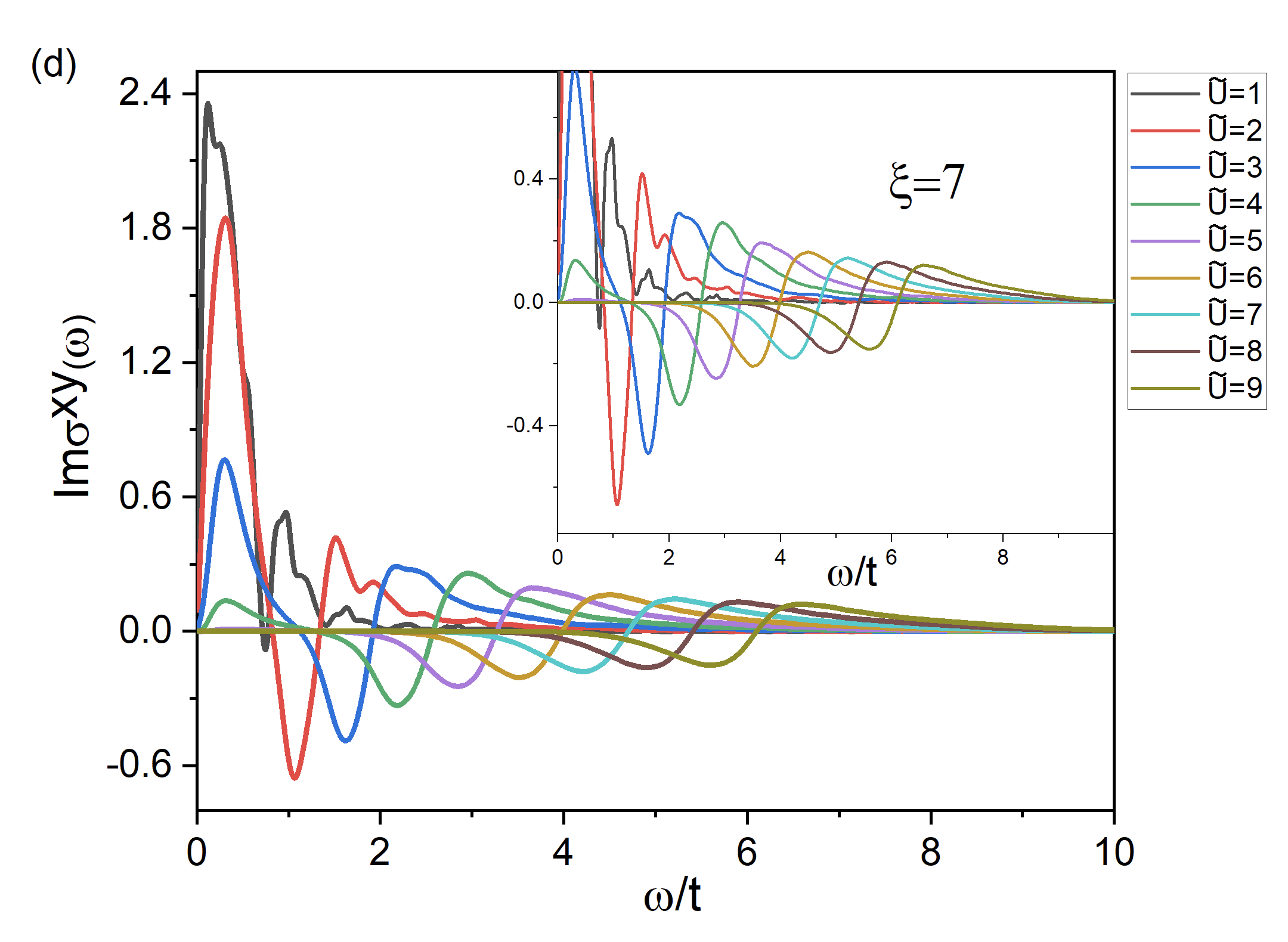}
\caption{The imaginary part of the infrared Hall conductivity of the system when we assume a non-Gaussian action with the form of the NLSM for the thermally fluctuating local moment. Such an action describes the thermal fluctuation in the renormalized classical regime of a quantum Heisenberg antiferromagnetic model defined on the square lattice. It is thought to be relevant to the electron-doped cuprate superconductors. The $\delta$-function peak is broadened into a Lorentzian peak of width $0.03t$. The calculation is done on a finite cluster of square lattice with $L\times L=400$ sites and periodic boundary condition in both the $x$ and the $y$-direction. The inset represents a zoom in view of the Hall conductivity spectrum. We note that the tiny negative overshoot near $\omega=0$ can be attributed to finite size effect in the anti-symmetrization procedure.}
\end{figure}

 \begin{figure}
\includegraphics[width=7cm]{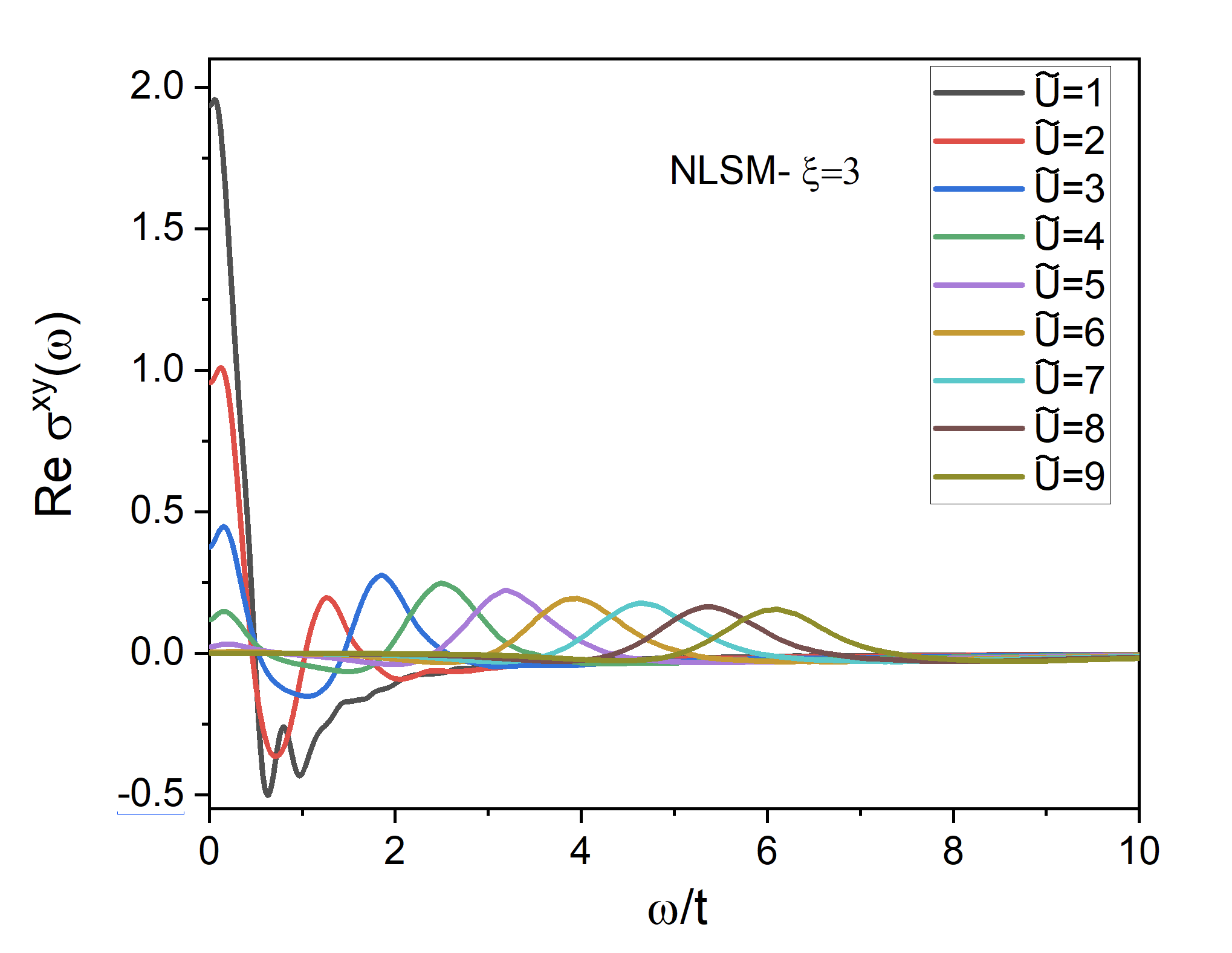}
\caption{The real part of the infrared Hall conductivity of the system when we assume a non-Gaussian action with the form of the NLSM and $\xi=3$ for the thermally fluctuating local moment. The calculation is done on a finite cluster of square lattice with $L\times L=400$ sites and periodic boundary condition in both the $x$ and the $y$-direction..}
\end{figure}  

\subsection{The optical and Hall conductivity in an antiferromagnetic ordered state}
To understand the numerical results presented in the last subsection, we compute the optical conductivity and the Hall conductivity in the antiferromagnetic long-range ordered state with the mean field approximation. This amounts to set
\begin{equation}
\bm{\phi}_{i}=\mathbf{m}e^{i\mathbf{Q}\cdot \mathbf{r}_{i}}
\end{equation}
in which $\mathbf{Q}=(\pi,\pi)$ denotes the wave vector of the antiferromagnetic order. $\mathbf{m}$ denotes an arbitrary vector. Here we assume $\mathbf{m}$ to be directed in the $z$-direction and denote its magnitude as $m$. The mean field Hamiltonian of the electron then reads
\begin{equation}
H_{MF}=\sum_{\mathbf{k},\sigma}\epsilon_{\mathbf{k}}c^{\dagger}_{\mathbf{k},\sigma}c_{\mathbf{k},\sigma}-\frac{2Um}{3}\sum_{\mathbf{k},\sigma}\sigma c^{\dagger}_{\mathbf{k+Q},\sigma}c_{\mathbf{k},\sigma}
\end{equation}
in which 
\begin{equation}
\epsilon_{\mathbf{k}}=-2t(\cos k_{x}+\cos k_{y})-4t'\cos k_{x} \cos k_{y}-\mu
\end{equation}
is the bare electron dispersion. Denoting
\begin{equation}
\Psi_{\mathbf{k},\sigma}=\left(\begin{array}{c} c_{\mathbf{k},\sigma} \\ c_{\mathbf{k+Q},\sigma}\end{array}\right)
\end{equation}
and $\Delta_{\mathrm{AF}}=\frac{2Um}{3}$, the mean field Hamiltonian Eq.80 can be recasted in the form of  
\begin{equation}
H_{MF}=\sum_{\mathbf{k}\in\mathrm{AFBZ},\sigma}\Psi^{\dagger}_{\mathbf{k},\sigma}\left(\begin{array}{cc} \epsilon_{\mathbf{k}} & -\sigma\Delta_{\mathrm{AF}} \\-\sigma\Delta_{\mathrm{AF}} & \epsilon_{\mathbf{k+Q}}\end{array}\right)\Psi_{\mathbf{k},\sigma}
\end{equation}
here $\mathrm{AFBZ}$ denotes the antiferromagnetic Brillouin zone(as illustrated by the red rhombus in Fig.6). $H_{MF}$ can be easily diagonalized, with the result
\begin{equation}
H_{MF}=\sum_{\mathbf{k}\in\mathrm{AFBZ},s=\pm,\sigma}E_{\mathbf{k},s}\gamma^{\dagger}_{\mathbf{k},s,\sigma}\gamma_{\mathbf{k},s,\sigma}
\end{equation}
Here
\begin{equation}
E_{\mathbf{k},\pm}=\frac{\epsilon_{\mathbf{k}}+\epsilon_{\mathbf{k+Q}}}{2}\pm\frac{\sqrt{(\epsilon_{\mathbf{k}}-\epsilon_{\mathbf{k+Q}})^2+4\Delta_{\mathrm{AF}}^2}}{2}
\end{equation}
and 
\begin{equation}
\left(\begin{array}{c} c_{\mathbf{k},\sigma} \\ c_{\mathbf{k+Q},\sigma}\end{array}\right)=\left(\begin{array}{cc}u_{\mathbf{k},\sigma} & -v_{\mathbf{k},\sigma} \\v_{\mathbf{k},\sigma} & u_{\mathbf{k},\sigma}\end{array}\right)\left(\begin{array}{c} \gamma_{\mathbf{k},-,\sigma} \\ \gamma_{\mathbf{k},+,\sigma}\end{array}\right)
\end{equation}
with
\begin{eqnarray}
u_{\mathbf{k},\sigma}&=&\frac{\sigma\Delta_{\mathrm{AF}}}{\sqrt{(\epsilon_{\mathbf{k}}-E_{\mathbf{k},-})^{2}+\Delta_{\mathrm{AF}}^{2}}} \nonumber\\
v_{\mathbf{k},\sigma}&=&\frac{\epsilon_{\mathbf{k}}-E_{\mathbf{k},-}}{\sqrt{(\epsilon_{\mathbf{k}}-E_{\mathbf{k},-})^{2}+\Delta^{2}}} \nonumber\\
\end{eqnarray}

In the antiferromagnetic ordered state, the fermi surface of the system will be reconstructed. As is illustrated in Fig.6, the non-interacting system has a large hole fermi surface around $(\pi,\pi)$. When we turn on the antiferromagnetic order, the unit cell will be doubled and the fermi surface will be folded by $\mathbf{Q}$. The reconstructed fermi surface is composed of four hole pockets around $(\pm\pi/2,\pm\pi/2)$ and two electron pockets around $(0,\pi)$ and $(\pi,0)$. For our choice of parameters, the hole pocket dominates the electron pocket. This has important consequence on the DC Hall response of the system.

\begin{figure}
\includegraphics[width=7cm]{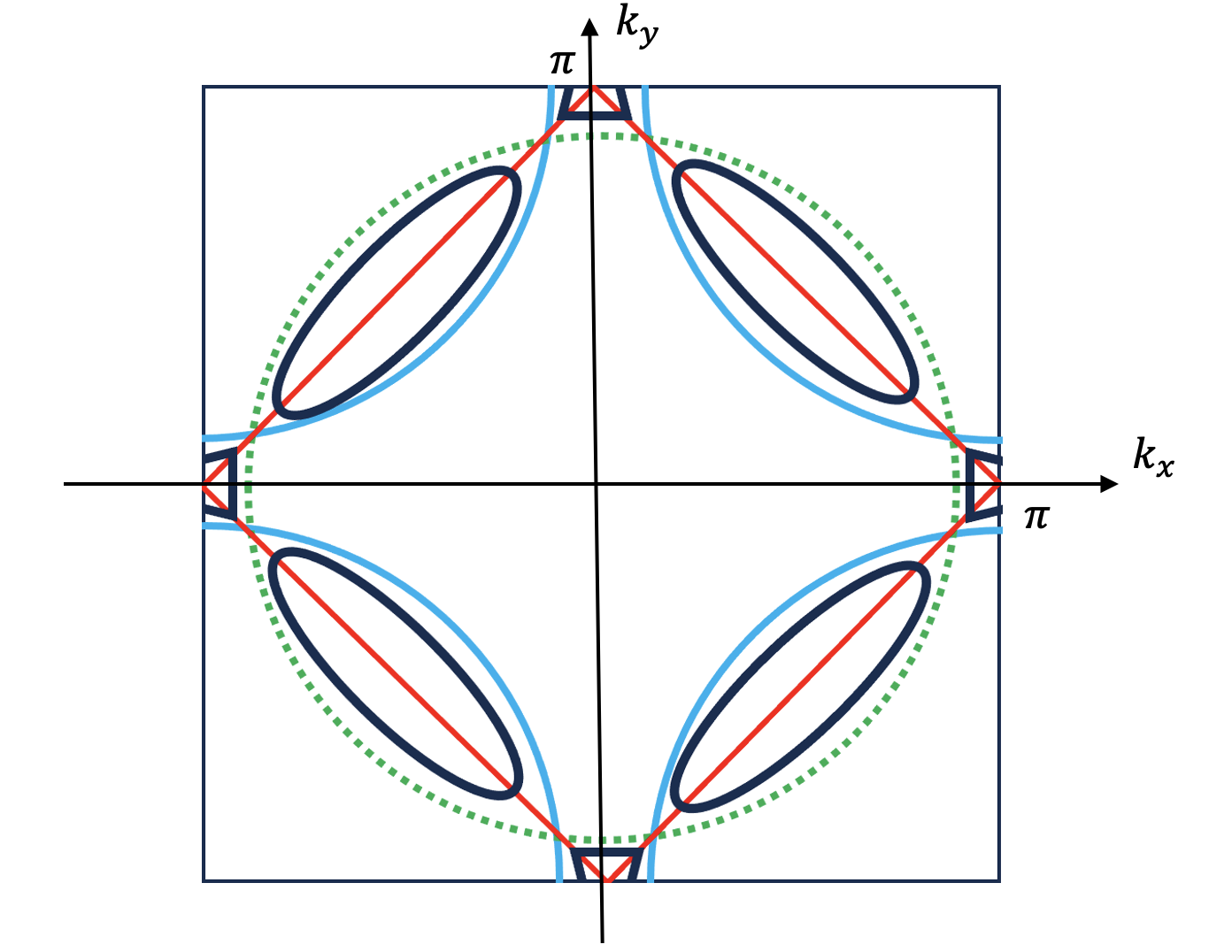}
\caption{Illustration of the fermi surface of the system considered. Here we set $t'=-0.3t$ and $\mu=-t$. The thick blue line marks the fermi surface of the non-interacting system. The red rhombus denotes the boundary of the reduced Brillouin zone of the antiferromagnetic ordered state. The green dotted line denotes the antiferromagnetic folded fermi surface. The thick black lines denote the reconstructed fermi surface in the antiferromagnetic ordered state. It is composed of four hole pockets around $(\pm\pi/2,\pm\pi/2)$ and two electron pockets around $(\pi,0)$ and $(0,\pi)$. For our choice of parameters the hole pocket clearly dominates the electron pocket.}
\end{figure}  

With these preparations, we can compute the optical conductivity and the Hall conductivity of the system from Eq.59 and Eq.62. The summation over the single particle eigenstate index $m$ appearing in Eq.63-66 now reduces to the summation over the momentum $\mathbf{k}$ in the AFBZ and that over the SDW band index $s$. More specifically, we have
\begin{eqnarray}
\sigma^{xx}(\omega)&=& \frac{\pi}{\omega}\sum_{\mathbf{k}\in AFBZ}\sum_{s,s'=\pm}\left|[\mathrm{v}^{x}_{\mathbf{q}}]_{\mathbf{k+q},s',\mathbf{k},s}\right|^{2}\nonumber\\
&\times&[f(E_{\mathbf{k+q},s'})-f(E_{\mathbf{k},s})]\nonumber\\
&\times&\delta(\omega-[E_{\mathbf{k},s}-E_{\mathbf{k+q},s'}])|_{|\mathbf{q}|\rightarrow 0} 
\end{eqnarray}
We note that a small but nonzero wave vector must be introduced in the calculation of the optical conductivity in the mean field approximation. In the mean field approximation, the electron momentum is strictly conserved up to $\mathbf{Q}$. The lack of momentum relaxation in the mean field approximation makes the intra-band transition forbidden at $\mathbf{q}=0$. A small but nonzero wave vector is thus necessary to recover the correct intra-band optical spectral weight.

The calculated optical conductivity with a SDW gap $\Delta_{\mathrm{AF}}=0.3t$ is shown in Fig.7a. The two-component structure is clearly seen. Obviously, the Drude peak at low energy should be attributed to the intra-band transition and that the mid-infrared peak should be attributed to the transition across the SDW gap. Indeed, the mid-infrared peak starts at an energy of approximately $2\Delta_{\mathrm{AF}}$. Thus, the two-component structure we observed in the optical conductivity in the last subsection can be understood as the precursor effect of the antiferromagnetic long range order. Indeed, the optical conductivity in Fig.1 and Fig.2 in the last subsection can be viewed as a smeared version of Fig.7a. 

 \begin{figure}
\includegraphics[width=7cm]{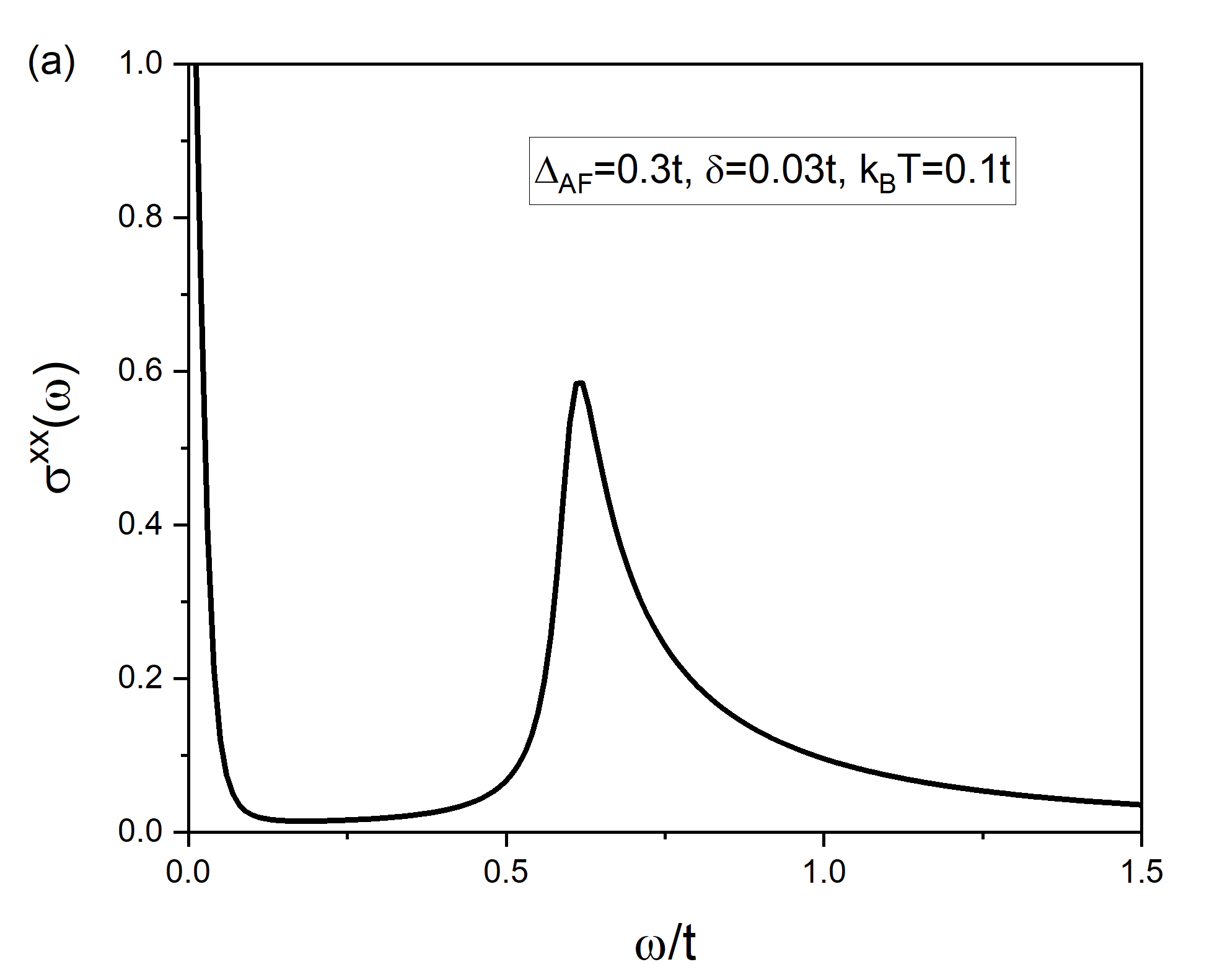}
\includegraphics[width=7cm]{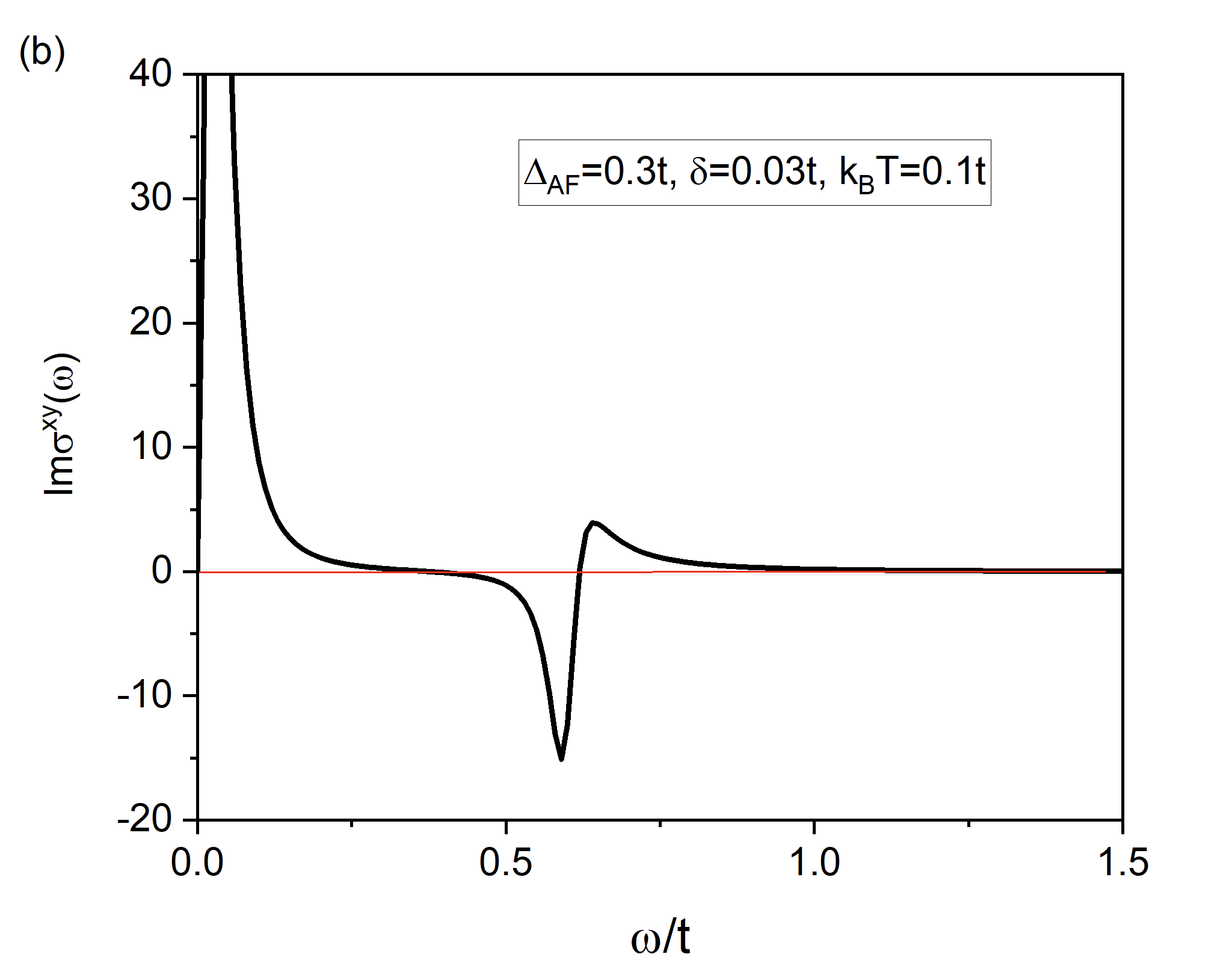}
\caption{The optical conductivity and the Hall conductivity spectrum of the system when we assume a static antiferromagnetic long range order and a SDW gap of $\Delta_{\mathrm{AF}}=0.3t$. The energy is measured in unit of $t$ and we have set $k_{B}T=0.1t$ for the electron. The calculation is done on a $L\times L=10^{6}$ cluster and the $\delta$-function peak is broadened into a Lorentzian peak of width $0.03t$.}
\end{figure}

The same consideration applies to the Hall conductivity. The calculated Hall conductivity spectrum $\mathrm{Im}\sigma^{xy}(\omega)$ with $\Delta_{\mathrm{AF}}=0.3t$ is shown in Fig.7b. The two-component structure and the accompanying sign changes between the low energy peak and the mid-infrared peak can be clearly seen. In fact, the calculated Hall conductivity spectrum shown in Fig.3 and Fig.4 in the last subsection can be viewed as a smeared version of Fig.7b. Clearly, the low energy peak should again be attributed to the intra-band transition and that the mid-infrared peak should be attributed to the inter-band transition between the SDW-split bands. We note that for our choice of parameters, the hole pocket always dominates the electron pocket on the reconstructed fermi surface. We thus expect that the overall contribution of the intra-band transition to the Hall conductivity spectrum to be positive. Indeed, the low energy peak in the calculated $\mathrm{Im}\sigma^{xy}(\omega)$ is always positive. This is different from the situation in Ref.[\onlinecite{Lin}], in which the electron pocket dominates the hole pocket. Correspondingly, while the low energy peak is positive when $\Delta_{\mathrm{AF}}$ is small, it can be driven negative when $\Delta_{\mathrm{AF}}$ is sufficiently large. In that case, there will be only one sign change between the low energy peak and the mid-infrared peak. 

According to such a scenario, the sign change in $\mathrm{Im}\sigma^{xy}(\omega)$ below the mid-infrared peak should be understood as the precursor of the fermi surface reconstruction and more specifically, the emergence of electron pocket which contribute negatively to the Hall response in the DC limit. We note that such a sign change has indeed been observed in the electron-doped cuprate superconductors. This implies that the antiferromagnetic correlation is indeed playing a crucial role in shaping the Hall response of the system. The discussion above also implies that the Hall response in the low frequency limit is a very subtle and may depends sensitively on the details of the low energy physics of the system.

\section{Conclusions and discussions}
In this work, we have derived exact formulas for the EM response of the two dimensional Hubbard model in terms of an effective theory description of its low energy physics. We find that the optical conductivity and the Hall conductivity of the system can be represented as the ensemble average of the EM response kernel of a non-interacting system coupled to fluctuating moment. We also find that the notorious negative sign problem in the Monte Carlo sampling of such fluctuating moment disappears in two important limits, namely the high temperature limit in which thermal fluctuation dominates and the weakly correlated limit in which the perturbative series in $U$ converges.

It is interesting to note that the occurrence of the negative sign problem is deeply related to the divergence of the perturbative series, which is again closely related to the emergence of local moment in the low energy physics of the system. This implies the possibility of solving the negative sign problem by finding a sign-problem-free approximation for the action of the fluctuating local moment. We find that the conventional $\phi^{4}$ approximation with a local interaction term just fulfills such a need. In particular, we show that the Monte Carlo simulation of local moment fluctuation is free from the negative sign problem if we assume the widely used MMP-type susceptibility in the effective action for the local moment, even though it is Landau damped as a result of its coupling to the itinerant quasiparticle on the fermi surface. We thus think that the combination of the low energy effective theory consideration and the standard quantum Monte Carlo simulation technique can be very fruitful in the numerical study of strongly correlated electron system like the cuprate superconductors. 

We have applied our formulas to study the effect of the thermal fluctuation of the local moment on the optical conductivity and the Hall conductivity of the cuprate superconductors. Using a Gaussian effective action with a MMP-type susceptibility for the fluctuating local moment, we have simulated how the optical conductivity and Hall conductivity evolve with the correlation length $\xi$ and the coupling strength with the local moment $\tilde{U}$. The optical conductivity calculated from our numerical simulation is found to exhibit a two-component structure, with a Drude component at low energy and a mid-infrared component at higher energy. Such a two-component characteristic is found to become more evident as we increase the correlation length and the coupling strength. In fact, we find that these two components can be understood as the remnant of the intra-band and inter-band optical weight in an antiferromagnetic ordered state, in which the electron band is split into the upper and the lower SDW band. 

We note that while such a two-component behavior is ubiquitously reported in the optical spectrum of the cuprate superconductors, the Drude weight predicted here is too weak to be consistent with the experimental observations in the strong coupling regime. We think that such a discrepancy should be attributed to the neglect of quantum nature of the local moment fluctuation when we are dealing with the EM response of the system in the low frequency regime. More specifically, at time scale much longer than that of the local moment fluctuation, it is expected that the electron density of state around the fermi level should be partially recovered from the SDW gapping by the thermal spin fluctuation. Such a quantum effect is expected to be more important in the hole-doped cuprate superconductors than in the electron-doped cuprate superconductors, since the local moment fluctuation is more dynamic in the former. We note that an exact treatment of such quantum effect is also possible with sign-problem-free quantum Monte Carlo simulation within our framework. A study of such an effect will be pursued in a future work. 

We find that the same two-component phenomenology also applies to the Hall conductivity spectrum $\mathrm{Im}\sigma^{xy}(\omega)$ of the system. The Drude component and the mid-infrared component in $\mathrm{Im}\sigma^{xy}(\omega)$ can be understood as derived from the corresponding spectral features caused by the intra-band and the inter-band transition in an antiferromagnetic long range ordered background. We find that for our choice of parameters both spectral components are positive and there exists two sign changes between them. Such sign changes can be taken as the precursor of the fermi surface reconstruction induced by the antiferromagnetic order, or more specifically, the emergence of electron pocket in such a state. We find that the Drude component in $\mathrm{Im}\sigma^{xy}(\omega)$ can be either positive or negative, depending on the relative importance of the hole pocket and the electron pocket on the reconstructed fermi surface. Thus, the Hall response in the low frequency limit is rather subtle and may depends sensitively on the details of the low energy physics of the system.

In addition, we find that the two-component characteristic in the optical conductivity and the Hall conductivity becomes more evident when we assume a non-Gaussian action of the form of the NLSM for the fluctuating local moment. Such an action describes the thermal spin fluctuation in the renormalized classical regime of a two dimensional quantum Heisenberg antiferromagnet defined on the square lattice and is thought to be relevant to the electron-doped cuprate superconductors. A semi-quantitative comparison of the result presented in Fig.2 and Fig.4 with the measured optical conductivity and the Hall conductivity spectrum in the electron-doped cuprate superconductors is thus very likely to be realistic.

The computational scheme developed in this paper is certainly not restricted to the studies presented in this work. First, our method can be equally well adopted to study the effect of quantum fluctuation of local moment. It is important to know to what extent the two-component structure we discovered in the optical conductivity and the Hall conductivity spectrum would survive when we take into account of the quantum nature of the local moment fluctuation. Experimentally, the more interesting but also more challenging quantities to be calculated are the longitudinal and Hall conductivity in the DC limit. As we argued above, the quantum nature of the local moment fluctuation plays a crucial role in the study of the DC transport properties. Second, while we focus on the EM response of two dimensional Hubbard model in this paper, our formalism can be naturally extended to study the effect of local moment fluctuation on other properties of such a strongly correlated system. For example, we can couple the electron to Grassmannian external field and calculate the single particle properties of the system with the current formalism. Within the current formalism, it is even possible to derive a self-consistent condition for the effective action of the fluctuating local moment by investigating the response of the system to external field applied in the spin channel. Third, the effective theory description is certainly not restricted to the study of local moment fluctuation. The MMP-type susceptibility can be naturally extended to describe other collective fluctuations, in particular those occurring in the charge density channel and the pairing channel for more general strongly correlated electron systems.

\begin{acknowledgments}
We acknowledge the support from the National Natural Science Foundation of China(Grant No.12274457). 
\end{acknowledgments}

\end{document}